\documentclass[amsmath,amsfonts,amssymb,twocolumn,superscriptaddress,showpacs]{revtex4}

\usepackage{graphicx,psfrag,color}% Include figure files
\usepackage{dcolumn}% Align table columns on decimal point
\usepackage{bm}% bold math
\usepackage{mathbbol}

\def\Z{{\mathbb{Z}}}
\def\tr{{\rm tr}\; }

\newcommand{\ket}[1]{| #1 \rangle}
\newcommand{\bra}[1]{\langle #1 |}
\newcommand{\proj}[1]{\ket{#1}\! \bra{#1}}

\begin{document}

\title{Autocorrelations and Thermal Fragility of Anyonic Loops \\ in
Topologically Quantum Ordered Systems}

\author{Zohar Nussinov}
\affiliation{Department of Physics, Washington University, St.
Louis, MO 63160, USA}
\author{Gerardo Ortiz}
\affiliation{Department of Physics, Indiana University, Bloomington,
IN 47405, USA}

\date{\today}

\begin{abstract}
Are systems that display Topological Quantum Order (TQO), and have a
gap to excitations, hardware fault-tolerant at finite temperatures?  We
show that in models that display low $d$-dimensional Gauge-Like
Symmetries, such as Kitaev's and its generalizations, the expectation
value of topological symmetry operators vanishes at {\it any} non-zero
temperature, a phenomenon that we coined {\it thermal fragility}. The
autocorrelation time for the non-local topological quantities in these
systems may remain finite even in the thermodynamic limit.  We provide
explicit expressions for the autocorrelation  functions in Kitaev's
Toric code model. If temperatures far below the gap may be achieved
then these autocorrelation times, albeit finite, can be made large. The
physical engine behind the loss of correlations at large spatial and/or
temporal distance  is the proliferation of topological defects at any
finite temperature as a result of a {\it dimensional reduction}. This
raises an important question:  How may we best quantify the degree of
protection of quantum information in a topologically ordered system at
finite temperature?
\end{abstract}

\maketitle

%%%%%%%%%%%%%%%%%%%%%%%%%%%%%%%%%%%%%%%%%%%%%%%%%%%%%%%%%%%%%%%%%%%%%
%%% Intro

\section{Introduction}

The perseverance of information over long times in the simplest
of memory devices is related to the existence of large associated 
autocorrelation times. The storage of information is intimately tied to
the breaking of ergodicity at scales much smaller than the autocorrelation 
time. Classical information can be reliably stored in magnetically or in
electrically (permanently) polarized materials. From the physicist's
perspective, this reliability is directly linked to the existence
of an {\it order parameter} (its {\it macroscopic} magnetization or
polarization) which characterizes a {\it collective} and robust
property of the material below an ordering transition temperature.
At its core, non-ergodicity implies the existence of 
a {\it generalized order parameter} (e.g. the overlap
parameter of spin glasses). 

The reliable storage of quantum information is a real
challenge. The uncontrolled interactions between a quantum system and
its environment or measurement apparatus introduce noise (errors) in
the system leading to decoherence of pure quantum superposition states.
Fortunately, quantum states can, in principle, be encoded fault
tolerantly and be {\it protected} against decoherence, thus preventing
loss of information \cite{Shor}. This idea lies at the heart of TQO
systems as first advanced by Kitaev \cite{kitaev}. Assuming that
errors are of a {\it local} nature, topological quantum memories (e.g.
surface codes \cite{kitaev}) seem to be intrinsically stable because of
{\it physical} fault-tolerance to weak quasi-local perturbations.
However, are these quantum memories robust to thermal effects?

In this work, we analyze the effect of temperature on zero-temperature
($T=0$) topologically ordered quantum systems \cite{wenbook,wen_plaq}, 
such as Kitaev's Toric Code \cite{kitaev} and Honeycomb models 
\cite{Kitaev2006} and generalizations thereof.  To this end we need to
present two concepts that were introduced for the first time in our
previous work \cite{NOLong}. One is the concept of finite-$T$
Topological Quantum Order (TQO), and the other of rank-$n$ TQO. In that
same work we studied the thermal fragility of topological operators in
$D=2$ lattice models. Our results \cite{NOLong} concerning the singular
character of the $T=0$ TQO in one notable system (Kitaev's Toric code
model) have later been reaffirmed in  work by Castelnovo and Chamon
\cite{CC} in their study of the topological entanglement entropy.  In
the present work we will present extensions of our ideas to higher
spatial dimensions $D$ and expand on the physical reasons leading to
thermal fragility. In particular, we  show that a general $\Z_{k}$
gauge theory in $D$ spatial dimensions in a system with periodic
boundary conditions displays rank-$n=k^{D}$ TQO. Nevertheless, although
a thermodynamic phase transition may occur, the system is thermally
fragile. We investigate not only the thermodynamic but also the
dynamical aspects of thermal fragility, and in cases such as Kitaev's
Toric Code model we also obtain exact analytic time-dependent results
thanks to our duality mappings \cite{NOLong}.

\section{Landau Orders vs TQO}
\label{sec2}

Before defining TQO, and to put this latter concept in perspective, let
us briefly review the rudiments of a  Landau order parameter. The
Landau order parameter is customarily associated with the breaking of a
global symmetry. The existence of an order parameter, a macroscopic
property measuring the degree of order in a state of matter, is
directly  associated to the phenomenon of spontaneous symmetry breaking
(SSB). This  concept, that involves an infinite number of degrees of
freedom, is so fundamental to condensed matter and particle physics
that many excellent textbooks (see, for example, \cite{anderson}) have
spent entire chapters (or even a full book  \cite{strocchi}) describing
it. For the present purposes, we illustrate the concept in the simple
case of a ferromagnet. A piece of iron at high temperatures it is in a
{\it disordered} paramagnetic phase. Below a certain temperature $T_c$
the system {\it orders}, i.e. it magnetizes,  and with the appearance
of the order parameter (magnetization) there is a breaking of the
rotational symmetry \cite{phase_trans}.  In the (ferro)magnetic phase
there is a net magnetization ${\bf M}$ that persists all the way to
zero temperature (where it attains its maximal value). The magnetization 
can, mathematically, be written as a linear
combination of {\it quasi-local} operators $V$ (e.g. $V$ is the local
spin operator in a Heisenberg model). The main point to stress here is
that the operator $V$ may distinguish between different ground states
(GSs) $\ket{g_\alpha}$ and $\ket{g_\beta}$ of the material 
\begin{eqnarray}
\langle g_{\alpha} | V | g_{\alpha} \rangle \neq \langle g_{\beta} | V |
g_{\beta} \rangle ,
\label{def.1}
\end{eqnarray}
and equivalently at finite temperatures, 
\begin{eqnarray}
\langle  V  \rangle_\alpha
\neq  \langle V \rangle_\beta.
\label{vab}
\end{eqnarray}
  Here and throughout 
the angular brackets refer to a thermodynamic 
average: for any quantiry- say ${\cal{A}}$ we have that 
$\langle \hat{\cal{A}} \rangle = \tr [ \rho \hat{\cal{A}}]$. 
The density matrix $\rho ={\cal Z}^{-1}
\exp[-H/(k_{B}T)]$ with ${\cal Z} = \tr [\exp[-\beta H]]$
where $\beta$ is the inverse temperature. 
Similarly, we define a (trace-class) density matrix $\rho_{\alpha} =
{\cal Z}_\alpha^{-1}\exp[-H_{\alpha}/(k_{B} T)]$ (with  ${\cal
Z}_\alpha = \tr [\exp[-\beta H_\alpha]]$ the partition function)
corresponding to the Hamiltonian $H$ endowed with terms which
favor order in the state $|g_{\alpha} \rangle$. These
subscripts ($\alpha$) are those appearing in the thermodynamic
averages of Eq.(\ref{vab}). A particular
realization of $\rho_{\alpha}$ for the problems
that will interest us will be given below 
(Eq.(\ref{rhoa})).

Symmetry plays a key role in dictating  the fundamental properties of
matter. Symmetry often generally implies the existence of conserved
charges with unique physical consequences. Most Landau orders are
inherently tied to broken global symmetries. There are symmetries other
than global. For instance, the symmetries in gauge theories are local;
such local symmetries cannot be broken \cite{Elitzur}.  Recently, a
general classification (and their physical consequences) of these and
other  types of symmetries was proposed \cite{NOLong}. A symmetry is
termed a  $d$-dimensional Gauge-Like-Symmetry ($d$-GLS) if the minimal
non-empty set of symmetry operations operate on a $d$-dimensional
spatial volume \cite{BN,NOLong}. Thus, global symmetries - those of the
usual Landau-type - correspond to $d=D$ (here the symmetry operators
act non-trivially on the entire $D$-dimensional system), and gauge
symmetries (which are local in nature) correspond to $d=0$ (as the 
symmetry operators act non-trivially only on quasi-local (or
$d=0$-dimensional) regions). General symmetries may lie anywhere in
between these two extremes: $0 \le d \le D$.  The groups associated
with such symmetries can be denoted  as ${\cal G}_d$ \cite{BN,NOLong}. 
(In the following we will only consider unitary representations.) The
statements that we will make below pertain to general systems, both in
the continuum and on lattices.  For explicit forms, in what follows, we
will often provide expressions and refer to  systems defined on
$D$-dimensional  hypercubic lattices of size $L \times L \times \cdots
\times L$. 

%{\it Topological Quantum Order: A Symmetry Principle.}
Let us now define TQO. Given a set of $n$ orthonormal GSs
$\{|g_{\alpha} \rangle\}_{\alpha=1,\cdots, n}$, with $1< n \le N_{g}$ 
where $N_{g}$ is the total number of GSs of a given Hamiltonian $H$,
$T=0$ rank-$n$ TQO exists iff for any bounded operator $V$ with compact
support (i.e.  any  {\it quasi-local operator} $V$),
\begin{eqnarray}
\langle g_{\alpha} | V | g_{\beta} \rangle = v \ \delta_{\alpha \beta} +
c_{\alpha \beta} ,
\label{def.}
\end{eqnarray}
where $v$ is a constant and $c_{\alpha \beta}$ is a correction that 
vanishes in the thermodynamic limit. This is indeed a condition on {\it
non-distinguishability} of GSs through local measurements.  Here and
throughout, we will employ Greek letters $\alpha$ and $\beta$ to denote
orthogonal states in the GS manifold. Note that Eq. (\ref{def.})
applies only to systems with degenerate GSs. Following standard
conventions, $\beta$ will also be employed for the inverse temperature
$1/(k_BT)$ where $k_B$ is the Boltzmann constant. 

General error detection in TQO systems, motivated by quantum error
detection conditions elsewhere, is given by
\begin{equation}
[\hat{P}_0 V \hat{P}_0, \hat{T}_\mu]= P_{0} [V,\hat{T}_{\mu}]P_{0}= 0,
\label{detection_cond}
\end{equation} 
where $\hat{P}_0=\sum_\alpha \proj{g_\alpha}$ is the protected
subspace, and $\hat{T}_\mu$'s represent the logical operators which are
not part of the {\it code's stabilizer} \cite{explain_stabilizer}. 
These operators are non-trivial symmetries of $H$, i.e.  
\begin{eqnarray}
[H,\hat{T}_{\mu}]=0 ,
\label{good}
\end{eqnarray} 
and encode the braiding operations that ensure topological degeneracy
of the GS manifold. In the anyonic schemes,
these operators represent braiding opeations.
We will reserve the use of the Greek indices  $\mu$
and $\nu$ to the operators $\{\hat{T}_{\mu}\}$. It is important to
emphasize that the $T=0$ TQO quantum error detection condition, Eq.
(\ref{detection_cond}), applies for systems with degenerate and
non-degenerate GSs (unlike Eq. (\ref{def.}) which only applies to
systems with degenerate GSs).

Clearly, when $c_{\alpha \beta} =0$  condition (\ref{def.}) implies
(\ref{detection_cond}). To see this, we write the commutator of Eq.
(\ref{detection_cond}) longhand to find that
\begin{eqnarray}
\sum_{\alpha, \beta} \langle g_{\alpha} | V| g_{\beta} \rangle
[|g_{\alpha} \rangle \langle g_{\beta}|, \hat{T}_{\mu}] = v
[\hat{P}_{0}, \hat{T}_{\mu}] = 0
\end{eqnarray}
identically (regardless of the specific symmetry operator 
$\hat{T}_{\mu}$). This follows from Eq. (\ref{good}) whenever Eq.
(\ref{def.}) holds with $c_{\alpha \neq \beta} =0$. Thus, condition 
(\ref{def.}) is sufficient (but not necessary) to ensure the general
error detection condition of Eq. (\ref{detection_cond}).

A finite-$T$ ($T >0$) generalization of TQO is provided by the
condition \cite{NOLong}
\begin{eqnarray}
\langle V \rangle_{\alpha}  \equiv \tr[\rho_{\alpha} V] &=& v 
+ c_{\alpha \alpha}(L) , \ \forall \alpha
\label{vt} 
\end{eqnarray} 
with $c_{\alpha \alpha}$ a correction that tends to zero in the
thermodynamic limit.  TQO systems
\cite{NOLong} satisfy both the $T=0$ as well as the finite-$T$
conditions of Eqs. (\ref{def.}) and (\ref{vt}).

Motivated by the conditions for quantum error detection one can propose
an extension of (\ref{detection_cond}) to finite temperatures. The finite
temperature error detection condition that we will focus on in this
work \cite{Manny} is
\begin{equation}
[\rho^{1/2} V \rho^{1/2}, \hat{T}_{\mu}]=
\rho^{1/2} [V, \hat{T}_\mu] \rho^{1/2}=0 ,
\label{detectionT_cond}
\end{equation} 
for all quasi-local operators $V$. Here, the propagation of a local
error at finite temperatures ($ V $) causes no harm to  the logical
operators $\{\hat{T}_{\mu}\}$. In the "typical finite temperature 
subspace", the local errors ($\tilde{V} = \rho^{1/2} V \rho^{1/2}$)
do not alter the algebra of the operators $\{\hat{T}_{\mu}\}$.
What we will ultimately measure at thermal equilibrium 
are objects of the form $\tr[\rho V_{1} \hat{T}_{\mu} V_{2} \hat{T}_{\nu}...]$.
[For finite times (in which equilibration has not set in
yet), we will measure finite time correlations of a similar form.]
If the appearance of the local operators $V_{i}$ does not alter
the algebra of the symmetry operators $\hat{T}_{\mu}$.
Eq.(\ref{detectionT_cond}) is the simplest caricature 
ensuring such invariances. In anyonic schemes, the algebra of such non-local
operators-- the algebra of $\{\hat{T}_{\mu}\}$-- is what enables quantum
memories. Similarly the algebra of related non-local operators encodes the
braiding operators that may perform topological quantum computing.
In the Appendix, we will show that Eq.
(\ref{detectionT_cond}) generally cannot be satisfied for any system of
a finite size nor, more generally, in any other system which  does not
display a finite-$T$ transition. The existence of a finite-$T$ phase
transition is a necessary but not sufficient condition for Eq.
(\ref{detectionT_cond}) to hold. We will now briefly relate a weaker
version of the finite-$T$ detection condition of Eq.
(\ref{detectionT_cond}) to  our earlier finite-$T$ TQO condition of Eq.
(\ref{vt}). This will suggest that although finite-$T$ transitions
(singularities in the free energy) are mandated to ensure finite 
temperature error detection, 
these transitions cannot be accompanied by SSB. Thus, these
transitions may be more akin to those in gauge theories. To this end, 
we note that if Eq. (\ref{detectionT_cond}) holds then, in particular,
the finite-$T$ expectation value 
\begin{eqnarray}
\langle V \rangle = \langle \hat{T}_{\mu}  V \hat{T}_{\mu}^{\dagger}
\rangle.
\label{VTV}
\end{eqnarray}
We now consider an extension of Eq. (\ref{VTV})  that is valid in the
thermodynamic limit for {\em all} quasi-local $V$
\begin{eqnarray}
\langle V \rangle_\alpha = \langle \hat{T}_{\mu}  V
\hat{T}_{\mu}^{\dagger} \rangle_\alpha ,
\label{VTVT}
\end{eqnarray}
stating that there is no SSB of the symmetries spanned by
$\{\hat{T}_{\mu}\}$ - at least insofar as any local observable $V$ can
detect. It is worth emphasizing that in Eq.(\ref{VTVT}),
the indices $\alpha$ and $\mu$ generally need not be the 
same. The absence of SSB detectable by local 
observables is the physical content of Eq. (\ref{vt}) with
\begin{eqnarray}
\rho_{\alpha} = {\cal Z}_\alpha^{-1}\exp[-\beta (H + h_\alpha
\hat{T}_{\alpha})],
\label{rhoa}
\end{eqnarray} 
where $h_\alpha \to 0$. 
%Thus, the finite-$T$ condition of  Eq. (\ref{vt}) is neccessary but
%not sufficient to ensure the requirement of Eq.
%(\ref{detectionT_cond}).

In our recent work \cite{NOLong}, we further  developed a symmetry
principle for TQO. We related certain symmetry transformations of a
system to the existence of TQO as defined by Eqs. (\ref{def.},
\ref{vt}). We emphasized the fundamental role $d$-GLSs play in
establishing that order and disallowing SSB of local quantities. We
basically proved sufficient symmetry conditions for a system to be
topologically quantum ordered: {\it When in a gapped system of finite
interaction range and strength, the GSs (each of which can be  chosen
by the application of an infinitesimal field) may be linked by 
discrete $d\leq 1$ or by continuous $d \le 2$ GLSs  $U \in
{\cal{G}_{\it d}}$, then a system that satisfies the $T=0$ conditions
of Eq. (\ref{def.}) exhibits finite-$T$ TQO [in the sense of Eqs.
(\ref{def.}, \ref{vt})].}  We refer the reader to \cite{NOLong} for a
comprehensive explanation.  The quantum error detection conditions of
Eq. (\ref{detectionT_cond}) are far more restrictive than the
finite-$T$ TQO conditions regarding the robustness of the system to all
quasi-local perturbations $V$.  
%We show in the Appendix that \ref{detectionT_cond}) 
%can only be satisfied in its zero temperature limit. 

Although it is possible in many cases \cite{NOLong} to satisfy the
weaker version of Eq. (\ref{detectionT_cond}) [that is Eq.
(\ref{VTVT})] by the use of $d$-GLSs, unless they are biased by hand,
the logical operators $\{\hat{T}_{\mu}\}$  always have a vanishing 
expectation value at any long-time equilibrium  finite-$T$ state (i.e.
any possible Gibbs state) in a system with finite range interactions
\begin{eqnarray}
\langle \hat{T}_{\mu} \rangle_\alpha =0.
\label{central}
\end{eqnarray}

A related, more practical,  consequence is that $\{\hat{T}_{\mu}\}$
{\em may  generally exhibit finite autocorrelation times}. That is, at
all positive temperatures,
\begin{eqnarray}
G_{\hat{T}_{\mu}}(t)= \langle \hat{T}_{\mu}(0) \hat{T}_{\mu}(t) \rangle 
\label{badmemory}
\end{eqnarray}
with $G_{\hat{T}_{\mu}}(t) \to 0$ as $(|t|/\tau) \to \infty$; there is
an inherently finite autocorrelation time $\tau$, whose size is limited
by thermal fluctuations (but not by system size). This autocorrelation
time remains finite even in the thermodynamic limit. An indefinitely
self-correcting TQO surface  code can only exist at exactly zero
temperature. However, it may be that by setting parameters
we can tune $\tau$ to be very large.

In Section \ref{d3kiv} we will show how this is  explicitly realized
in the Kitaev's Toric code model [Eqs. (\ref{gx5}, \ref{tauT},
\ref{tauTT})]. Before embarking on an analysis of specific cases, let
us first analyze general relations. 

In several specific cases, such as Kitaev's Toric code model, which  we
will analyze below, it is possible to find a new representation in
which an initial local Hamiltonian remains local yet {\em the 
non-local topological anyonic loops become objects of low effective
dimensionality}. For instance, in Kitaev's Toric code model, the
non-local Toric cycle loops become point $(d=0$) fields. In such cases
of low effective dimensionality, the absence of finite temperature SSB
in low dimensions guarantees that Eq. (\ref{central}) holds even
without performing more details expansions or bounds.  

Below, we prove this result for the particular case
of a symmetry operator which is independent not only of the code's
stabilizer but also of the Hamiltonian itself. Later on, we will
show how this follows also when the symmetry operator (or, in fact,
any non-local operator) is not
independent of the the arguments on which 
the Hamiltonian depends. When the symmetry operator is independent
of the Hamiltonian, a transformation
exists which turns the symmetry $\hat{T}_{\mu}$ into a $d=0$-dimensional 
operator. Similar to Eq.(\ref{rhoa}), we may define
\begin{eqnarray}
{\cal Z}(h_{\mu}) = \tr \Big[ \exp[-\beta(H-h_{\mu}\hat{T}_{\mu})] \Big]
\label{fullh}
\end{eqnarray}
with no summation over repeated indices ($\mu$) implicit.  If the
logical operator is independent of the argument of a local Hamiltonian
$H = \sum_{i} H_{i}$ (here $\{H_{i}\}$ are local operators), then  the
partition function of Eq. (\ref{fullh}) simply factorizes
\begin{eqnarray}
{\cal{Z}}(h_{\mu}) = {\cal{Z}} \times z(h_{\mu}).
\label{factorzz}
\end{eqnarray}
From Eq. (\ref{factorzz}), the expectation value
\begin{eqnarray}
\langle \hat{T}_{\mu} \rangle &=& \lim_{h_{\mu} \to 0^{+}} 
\frac{1}{\beta {\cal{Z}}(h_{\mu})} \frac{\partial}{\partial h_\mu}
{\cal{Z}}(h_{\mu}) \nonumber \\
 &=&  \lim_{h_{\mu} \to 0^{+}} \frac{1}{\beta
z(h_{\mu})}\frac{\partial}{\partial h_\mu}z(h_{\mu})=0.
\label{zexp}
\end{eqnarray}
The expectation value evaluated with $z(h_{\mu})$ that encompasses only
one site, cannot exhibit SSB (in formal terms, $z(h_{\mu})$ is the
partition function of a $d=0$-dimensional system). Thus, quite
universally, the expectation value of any such logical operator
vanishes. We emphasize that this holds for all systems (both in the
thermodynamic limit and finite size systems). All that matters in the
derivation of Eqs. (\ref{factorzz}, \ref{zexp}) is that $\hat{T}_{\mu}$
is independent of the variables $\{H_{i}\}$ that are added to form the
code's Hamiltonian. In Section \ref{sec3}  we will work out these and
related expectation values  in detail for Kitaev's Toric code model.

Equation (\ref{central}) may be extended more generally to {\it
non-local} high-dimensional operators ($\hat{R}_{a}$) that need not (i)
lie outside the code's stabilizer or (ii) be symmetries of $H$.  For
instance, in lattice gauge theories, $\hat{R}_{a}$ can pertain to a
Wilson loop of a divergent perimeter. \cite{explain_14.5} By
performing a low temperature series expansion about the ordered state,
one generally finds that, similar to Eq. (\ref{central}),   all
non-local operators $\{\hat{R}_{a}\}$ of dimension $d>0$ have a
vanishing expectation value at finite temperatures,
\begin{eqnarray}
\langle \hat{R}_{a} \rangle = 0.
\label{central+}
\end{eqnarray}
Equation (\ref{central+}) follows from an asymptotic perimeter law type
bound (see, e.g. \cite{kogut})
\begin{eqnarray}
|\langle \hat{R}_{a} \rangle| \le  A e^{-cm} ,
\label{perimeter}
\end{eqnarray}
with $m = {\cal{O}}(L^{d})$  the number of local fields that lie in the
support of $\hat{R}_{a}$ and $A, c$ positive constants. \cite{explain_15.5}
As seen from Eq. (\ref{perimeter}), for $m \to \infty$ (as befits any
non-local $\hat{R}_{a}$), this expectation value vanishes.  Hand in
hand, 
\begin{eqnarray}
\langle \hat{R}_{a}(0) \hat{R}_{a}(t) \rangle_{|t| \to \infty} \to 0.
\end{eqnarray}
Related results (for both the commutator in Eq. (\ref{detectionT_cond})
as well as the commutator between topological quantities  are afforded
by  simple extensions (carried in the Appendix) of the Lieb-Robinson
bounds known to apply for local  quantities in spin systems with local
interactions \cite{LR}. 

Consider next $d=0$ symmetries, i.e. operators $\{\hat{T}_{\mu}\}$ that
span only a zero-dimensional volume (or a finite number of points on a
lattice). For the operators $\{\hat{T}_{\mu}\}$ to realize a
non-trivial ray representation which leads to a topological
degeneracy,  these local symmetry operators cannot commute with one
another. Elitzur's theorem \cite{Elitzur} states that any quantity
which is not invariant under all local symmetries must vanish at any
finite temperature. Thus, for any quasi-local $\hat{T}_{\mu}$ symmetry 
operator (including all operators which may be defined on any finite
size lattice),  there can never be a SSB of $\hat{T}_{\mu}$ and once
again Eq. (\ref{central}) follows. Considerations 
similar to those of Eq.(\ref{zexp}) can be enacted. \cite{explain_15.75} 
Couched in the language more commonly used by researchers
in anyonic quantum computing, we can say that in this case- the
case of general local ($d=0$) operators (which includes any 
{\it finite} lattice as a special realization of $d=0$
symmetries which are here enforced by the limited physical extent $D=0$
of the system)- if we given a set of symmetry group operators 
$\{\hat{T}_\mu\}$, i.e. $[H,\hat{ T}_\mu]=0$, such that they form a
non-Abelian group or an Abelian group with a ray (non-vector)
representation, then we can prove that $\langle \hat{ T}_\mu
\rangle = 0$.  For example, in the case of a ray representation $\hat{
T}_\mu \hat{ T}_\nu =e^{i\phi_{\mu\nu}} \hat{T}_\nu \hat{ T}_\mu$,
$\langle \hat{ T}_\mu \rangle = \tr[\hat{T}_\nu \rho
\hat{T}^\dagger_\nu \hat{T}_\mu]= e^{i\phi_{\mu\nu}} \langle \hat{
T}_\mu \rangle=0$. From the dimension of the irreducible representation
we infer the degeneracy of the GS subspace. (We remind the reader that
for a continuous connected group of symmetries all finite-dimensional
ray representations are equivalent to vector representations.)

We conclude this section with a general remark and reiterate 
one of our
earlier comments regarding autocorrelation times:
Although indefinite quantum error
detection may be ruled out by a system that violates the simplest
finite T error detection condition of Eq.(\ref{detectionT_cond})) 
or for which Eq.(\ref{central}) holds, our results
do not rule out quantum error correction over
time scales that can be made quite large
(albeit still finite) by a judicial choice 
of parameters. In the next section,
we will show how although Eq.(\ref{central})
is realized in thermal equilibrium, the autocorrelation
time may be made large at very low temperatures.

\section{Thermal Fragility: worked-out examples}
\label{sec3}

In earlier work \cite{kitaev},  the presence of a gap in the energy
spectrum together with the existence of TQO as defined above (Eq.
(\ref{def.}) with $c_{\alpha\beta}=0$) were suggested  to be sufficient
to guarantee the protection of quantum information. The physical
intuition behind this was that properties of the protected (GS)
subspace are stable with respect to weak local perturbations. How does
temperature affect this conclusion? Is there any other additional
requirement needed for protection?  We just mentioned that the simple
generalization of the quantum error detection condition, Eq.
(\ref{detectionT_cond}), may generally fail at all temperatures $T \neq
0$. It is commonly believed that the existence of a finite gap between
the ground and first excited states protects properties associated with
$T=0$ TQO up to a finite energy scale $k_B T$ smaller than the gap
$\Delta$ since thermal fluctuations are suppressed by the Boltzmann
factor $\exp[-\Delta/k_BT]$. Our results \cite{NOLong} showed that
this  assumption is, in general,  incorrect for long times. 
We originally coined the
term {\it thermal fragility} to describe this state of affairs
\cite{NOLong}. We showed that in some prominent TQO models,  the entropic
weight associated with defects outweights their Boltzmann penalty of
$\exp[-\Delta/k_BT]$  at any finite temperature: the equilibrium 
states are always disordered. In what follows, we
expand on these concepts, determine the equilibration
time in one solvable case, and consider extensions of previous examples
to higher spatial dimensions $D$. In Section \ref{d3kiv}, we derive the
autocorrelation function for Kitaev's Toric code model. We then review
[in Section \ref{Knew}] a new high-dimensional extension of Kitaev's
Toric code model and prove that this system displays TQO and a finite
autocorrelation time. Finally, in Section \ref{kith} we discuss the
situation for Kitaev's Honeycomb Model.

\subsection{$D=2$ Kitaev's Toric Code Model}
\label{d3kiv}

\begin{figure}
\centerline{\includegraphics[width=0.9\columnwidth]{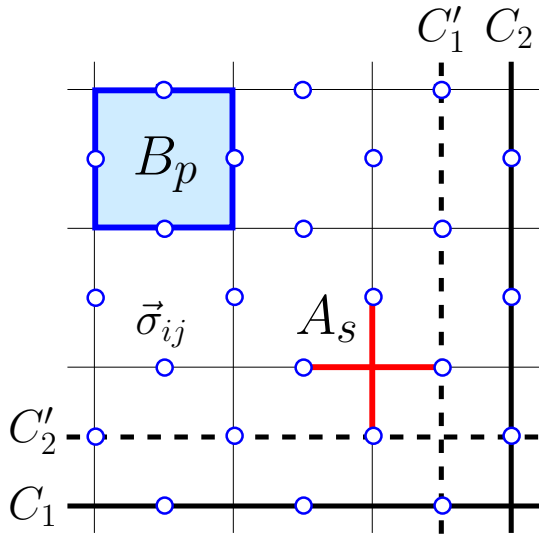}}
\caption{Elementary (cell) plaquette $B_p$ and star $A_s$ interaction
terms in Kitaev's Toric Code model. Hollow circles in the bonds (links)
represent an  $S=1/2$ degree of freedom, while thick (dashed or solid)
lines represent the topological symmetry operators (see text).}
\label{2dKitaev}
\end{figure}

For the sake of clarity and because of historical importance we start
by summarizing the main findings of Ref. \cite{NOLong} regarding
Kitaev's Toric Code model \cite{kitaev} in $D=2$. The model is defined
on a square lattice with $L \times L=N_s$ sites, where on each bond (or
link) $(ij)$ it is defined a $S=1/2$ degree of freedom indicated by a
Pauli matrix $\vec{\sigma}_{ij}$, thus defining a Hilbert space of
dimension $2^{2N_s}$ (see Fig.  \ref{2dKitaev}). The model Hamiltonian 
is given by 
\begin{eqnarray}
H_{K} = -\sum_{s} A_{s} -\sum_{p} B_{p}
\label{kitaevmodel}
\end{eqnarray}
with Hermitian operators (whose eigenvalues are $\pm 1$)
\begin{eqnarray}
A_{s} = \prod_{(ij) \in {\sf star}(s)} \sigma_{ij}^{x}, 
~~B_{p} = \prod_{(ij) \in {\sf plaquette}(p)} \sigma_{ij}^{z} ,
\label{AB_defn}
\end{eqnarray}
and $\sigma_{ij}^{\kappa}$ ($\kappa=x,y,z$) representing Pauli
matrices. $B_{p}$ and $A_{s}$ describe the plaquette (or face) and star
(or vertex) operators, respectively, with ($\forall s,s',p,p'$)
\begin{eqnarray}
[A_s,A_{s'}]=[B_p,B_{p'}]=[A_s,B_p]=0 ,
\end{eqnarray} 
thus generating an Abelian group called {\it code's stabilizer}
\cite{kitaev}. In the presence of periodic boundary conditions, the
plaquette and star operators satisfy the constraint
\begin{eqnarray}
\prod_{s} A_{s} = \prod_{p} B_{p} =1 ,
\label{ABpc}
\end{eqnarray}
and the two $d=1$ $\Z_{2}$ symmetries are given by  \cite{kitaev}
\begin{eqnarray}
Z_{1,2} = \prod_{(ij) \in C_{1,2}} \sigma_{ij}^{z}, ~X_{1,2} =
\prod_{(ij) \in C^{\prime}_{1,2}} \sigma^{x}_{ij},  \nonumber \\ 
\{X_{\mu}, Z_{\mu} \} = 0 \ , \ [X_{\mu}, Z_{\nu}] = 0 \  , \ \mu \neq
\nu ,
\label{kit}
\end{eqnarray}
where $C_1(C_2')$ are horizontal and $C_2(C_1')$ vertical closed
contours (i.e. loops on the lattice(dual lattice)).  The logical
operators $Z_{1,2}$ and $X_{1,2}$ commute with the code's stabilizer
but are not part of it, thus acting non-trivially on the two {\it
encoded} Toric code qubits. 

As shown in Ref. \cite{NOLong} $H_K$ is related to Wen's plaquette
model \cite{wen_plaq} {\it and} to two Ising chains by {\it exact}
duality mappings. Therefore, these three models share the {\it same}
spectrum. The GS (protected subspace of the code) is 4-fold degenerate
(Abelian $\Z_2\times \Z_2$  symmetry) and there is a gap to
excitations. The spectrum is basically that of two uncoupled circular
Ising chains ($2N_s$ is the total number of links of the original $D=2$
lattice) 
\begin{eqnarray}
\tilde{H}_K=-\sum_{p=1}^{N_s} \sigma^z_p \sigma^z_{p+1}
-\sum_{s=1}^{N_s} \sigma^x_s \sigma^x_{s+1} ,
\label{thG}
\end{eqnarray}
with GS energy $E_0=-2N_s$ and a gap to the first excited state equals
to 4 (this value of 4 and not 2 arises as in the presence of periodic
boundary conditions, only an even number of domain walls are possible).
This example clearly illustrates the fact that TQO is a property of
states and not of the Hamiltonian spectrum \cite{NOLong}.

The elementary excitations of $H_K$ are of two types \cite{kitaev}
\begin{eqnarray}
\ket{\Psi_z(\Gamma)}=\prod_{(ij)\in\Gamma} \sigma_{ij}^z \ket{\Psi_0}
\equiv S^{z}(\Gamma) \ket{\Psi_0}, \nonumber \\  
\ket{\Psi_x(\Gamma')}=\prod_{(ij)\in\Gamma'} \sigma_{ij}^x \ket{\Psi_0}
\equiv S^{x}(\Gamma') \ket{\Psi_0} ,
\label{QQbar}
\end{eqnarray}
where $\Gamma(\Gamma')$ is an open string on the lattice(dual lattice)
and $\ket{\Psi_0}$ is a GS. (If $\Gamma(\Gamma')$  would be closed
contours which circumscribe an entire Toric cycle then the string
operators $S^{x,z}$ would become the Toric symmetries of Eq.
(\ref{kit}).)  In the case of the open contours of Eq. (\ref{QQbar}),
the operators $S^{x,z}$ generate excitations at the  end points of
these strings (thus always coming in pairs) with Abelian  Fractional
Statistics (anyons). Excitations living on the vertices represent {\it
electric charges} while the ones living on the plaquettes are {\it
magnetic vortices}. These magnetic and electric type excitations obey
fusion rules that enable Abelian quantum computation.   Due to the
exact equivalence between Kitaev's model and the Ising chains, no
non-trivial finite temperature SSB or other transitions can take place.
The spectrum exhibits a multitude of  low-energy states. At any finite
temperature, no matter how small, entropic contributions to the free
energy overwhelm energy penalties and lead to a free energy which is
everywhere analytic \cite{NOLong}. 

The operators of Eqs. (\ref{QQbar}) are not symmetries of  $H$ and, for
divergent loop size, are specific examples of the non-local operators
$\{\hat{R}_{a}\}$ that we considered earlier (see, e.g. Eq.
(\ref{perimeter})). In the one-dimensional Ising duality mapping of Eq.
(\ref{thG}), we may  represent the string operators of Eq.
(\ref{QQbar}) as the creation operators for  domain walls in the $D=1$
Ising model. When $\Gamma$ and $\Gamma'$ intersect at any even number
of bonds, a representation is 
\begin{eqnarray}
S^{z}(\Gamma) = \prod_{s_{1}<s \le s_{2}} \sigma^{z}_{s}, \ \ \ 
S^{x}(\Gamma') = \prod_{p_{1} < p \le p_{2}} \sigma^{x}_{p}.
\label{QQ}
\end{eqnarray} 
In Eq. (\ref{QQ}), $s_{1}$ and $s_{2}$ denote the endpoints of  the
string $\Gamma$ of Eq. (\ref{QQbar}). Similarly, $p_{1,2}$ are the
plaquettes which form the endpoints of the string  $\Gamma'$ on the
dual lattice.  When $\Gamma$ and $\Gamma'$ share an odd number of
bonds, we can represent the string operators as follows
\begin{eqnarray}
S^{z}(\Gamma) = \sigma^{z}_{p_{2}}
\prod_{s_{1}<s \le s_{2}} \sigma^{z}_{s} 
, \ \
S^{x}(\Gamma') = \prod_{p_{1} < p \le p_{2}} \sigma^{x}_{p}.
\label{QQo}
\end{eqnarray} 

For any {\it finite} lattice and $T>0$
\begin{eqnarray}
\langle Z_\mu \rangle&=& \frac{\tr[ \exp [ -\beta H_K] Z_\mu]}{\tr[\exp
[ -\beta H_K]]}=\frac{\tr [ X_\mu \exp [ -\beta H_K] X_\mu Z_\mu]}{\tr
[\exp [ -\beta H_K]]} \nonumber \\ 
&=& \frac{\tr [\exp [ -\beta H_K] X_\mu Z_\mu X_\mu]}{\tr [\exp [
-\beta H_K]]} =-\langle Z_\mu \rangle=0,
\end{eqnarray}
where we have used the property $[H_K,X_\mu]=0=[H_K,Z_\mu]$ and the
cyclic invariance of the trace $\tr$, which is performed over the
eigenstates of $\sigma^z_{ij}$ with eigenvalues $\pm 1$. Similarly,
$\langle X_\mu \rangle=0$. Indeed, this is a special case of the more
general argument depicted in Section \ref{sec2}.

This does not, in principle, preclude SSB in the thermodynamic limit.
One needs to restrict the configurations over which the trace is
performed. In the thermodynamic limit, derivatives of the partition
function and associated free energy need not be analytic single-valued
functions (when SSB occurs the expectation values depend on how the
limit $\vec{h} \to 0$ is taken). To this end, let us define the
(generating) partition function of the model with the constraint
(\ref{ABpc})  
\begin{eqnarray}
\hspace*{-3cm} {\cal Z} \!&=&  \!\!\tr \! \Big[\exp [ -\beta ( H_K -
\!\!\sum_{\mu=1,2}  (h_{x,\mu} X_{\mu}+h_{z,\mu} Z_{\mu}) ) ] \Big] \\
&=&\!\![ (2 \cosh \beta)^{N_{s}} + (2 \sinh \beta)^{N_{s}}]^{2} 
\cosh \beta h_{1} \cosh \beta h_{2} , \nonumber
\label{pfunction}
\end{eqnarray}
where $h_{\mu} = \sqrt{h_{x,\mu}^{2} + h_{z,\mu}^{2}}$. Thus, the free
energy  per bond, $F =  - \frac{\beta^{-1}}{2N_{s}} \ln {\cal
Z}[h_{\kappa,\mu}=0]$, is analytic for all finite $\beta$, and displays
a singularity at $T=0$ (inherited from the Ising chain). That means
that no finite-$T$ phase transition occurs in Kitaev's model. Moreover,
from Eq. (\ref{pfunction}), we can compute the expectation values of
the topological operators with the result
\begin{eqnarray}
\!\!\!\!\!\! \langle Z_{\mu} \rangle \! &=& \!\! \lim_{h_{z,\mu} \to 0^{+}}
\frac{\partial}{\partial (\beta h_{z,\mu})} \ln {\cal Z} = \!\! \lim_{h_{z,\mu}
\to 0^{+}}\!  \frac{h_{z,\mu}}{h_{\mu}} \tanh(\beta h_{\mu}) , \nonumber \\ 
\!\!\!\!\!\! \langle X_{\mu} \rangle \! &=& \!\! \lim_{h_{x,\mu} \to 0^{+}}
\frac{\partial}{\partial (\beta h_{x,\mu})} \ln {\cal Z} =\!\!  \lim_{h_{x,\mu}
\to 0^{+}}\!  \frac{h_{x,\mu}}{h_{\mu}} \tanh(\beta h_{\mu}) , \nonumber \\
&&\langle Z_{1} \rangle = \langle Z_{2} \rangle = \langle X_{1} \rangle 
= \langle X_{2} \rangle =0.
\label{expectpart}
\end{eqnarray}
This indicates that  {\em the existence of a gap in this system may not
protect a finite expectation value of the Toric code operators
$X_{1,2}$ or $Z_{1,2}$ for any finite temperature $T>0$}
\cite{NOLong,alicki}. At $T=0$ these expectation values are finite and
equal to unity, reflecting the non-analyticity of $F$ at $T=0$. The
physical reason behind this result is the proliferation of topological
defects (solitons) at any finite $T$. The Boltzmann suppression becomes
ineffective at sufficiently long times and 
this might be bad news for a robust quantum memory 
\cite{dennis}. In the presence of additional fields (see last terms in
Eq. (\ref{pfunction})), with the two component vector $\vec{h}_{\mu} =
(h_{x,\mu}, h_{z,\mu})$ at site $\mu$, we may define the two component
vector
\begin{eqnarray}
\hat{n}_{\mu} = \frac{1}{|\vec{h}_{\mu}|} (h_{x,\mu}, h_{z,\mu})
\end{eqnarray}
and set ($[H_K - \sum_{\mu=1,2} (h_{x,\mu} X_{\mu}+h_{z,\mu} Z_{\mu}),
Q_\mu ]=0$)
\begin{eqnarray}
Q_{1} &=& X_{1} n_{x1} + Z_{1} n_{z1}, \nonumber \\ 
Q_{2} &=& X_{2} n_{x2} + Z_{2} n_{z2}
\label{q1q2Ising}
\end{eqnarray} 
to be the counterparts of two single Ising spins $\sigma^{x}_{sc}$ and
$\sigma^{z}_{pd}$ which  are located at site numbers $c$ and $d$ of
the  two respective Ising chains (that of the $s$ and  that of the $p$
varieties). $c$ and $d$  can be chosen to be any integers such  that $1
\le c, d  \le N_{s} $. These two spins (along with the spins appearing
in Eq. (\ref{thG}) satisfy precisely the same algebra and set of
constraints as the original  variables in Kitaev's model [Eqs.
(\ref{kitaevmodel},  \ref{AB_defn}, \ref{kit})]. The vanishing
expectation values of Eq. (\ref{expectpart}) for both the finite and
infinite system can be understood as the statement that $\langle
\sigma^{z}_{sc} \rangle = \langle \sigma^{z}_{pd} \rangle =0$ for any
one-dimensional Ising system when the single on-site magnetic fields
$\vec{h}_{c,d} \to 0$. 

The entropy associated with $d=1$ type domain walls is logarithmic in
the system size $N_s$. By contrast, the energy penalty for these domain
walls is finite and size independent. As a result, for sufficiently
large systems, entropic gains will outweigh energy  penalties. In
particular, in the thermodynamic limit, there is no SSB of the $d=1$
GLSs at any temperature $T>0$. In \cite{NOLong}, we showed how 
all static correlation functions may be computed via our 
mapping to the Ising chain. 

We can now address the dynamical aspects of thermal fragility in
Kitaev's Toric code model. From its mapping to two uncoupled Ising
chains, we can immediately determine the time autocorrelation functions
of the Toric code operators. As we have shown, Kitaev's Toric code
operators $\{X_{1,2}, Z_{1,2}\}$ map into single spins in an  Ising
chain. Thus, we can  
employ the results obtained in \cite{Glauber,BP}
concerning autocorrelation of spins in Ising chains in a system with
Glauber-type dynamics. In the Appendices (Eqs. (\ref{master1},
\ref{master2})), we will outline 
standard master equations which used to
determine the dynamics. When these equations depend only on the
system's spectra (and not the precise real-space form of the GSs) then
we can use our mapping of Eq. (\ref{q1q2Ising}) to relate the dynamics 
of the non-local topological quantities of 
Kitaev's Toric code model to single 
spin dynamics in Ising chains. We 
find that
\begin{eqnarray}
G_{X_{\mu}}(t) \equiv \langle X_{\mu}(0) X_{\mu}(t) \rangle 
\end{eqnarray}
obeys, for $t>0$, the relation
\begin{eqnarray}
\frac{\partial G_{X_{\mu}}}{\partial t} = \chi \
\frac{\tanh^{2}\beta-1}{\tanh^{2} \beta +1} \ I_{0} \Big[t \tanh \beta 
\tanh 2 \beta \Big] \ e^{-\chi t} .
\end{eqnarray}
Here, $\chi$ is a constant setting the time scale for  the evolution of
the system and $I_{0}$ is  the modified Bessel function. For low
temperatures, at short times, \cite{BP}
\begin{eqnarray}
|t| \ll \chi^{-1},
\end{eqnarray}
the autocorrelation is given by 
\begin{eqnarray}
G_{X_{\mu}; \mbox{\footnotesize \sf short-time}}(t) \simeq e^{-\sqrt{2
[1- \tanh 2 \beta]} \chi |t|}.
\end{eqnarray}
At intermediate times, \cite{BP}
\begin{eqnarray}
\chi^{-1} \ll |t| \ll \chi^{-1} \frac{1}{1- \tanh 2 \beta}, 
\end{eqnarray}
the autocorrelation of the topological string operators is well
approximated by a Cole-Davidson  form \cite{CD} in the frequency domain
and stretched exponential in time, 
\begin{eqnarray}
G_{X_{\mu}; \mbox{\footnotesize \sf intermediate-time}}(t) \simeq
e^{-\sqrt{|t|/\tau}},
\label{gx5}
\end{eqnarray}
with the equilibration time
\begin{eqnarray}
\tau = \frac{\pi}{4 \chi (1- \tanh 2 \beta)}.
\label{tauT}
\end{eqnarray}
The constant $\tau$ is independent of the system size (for large
systems)  and is finite for all temperatures $T>0$. If the temperature
can be made much smaller than the gap, i.e. if $\beta \gg 1$, then
$\tau$, albeit being finite, can be made large ($ \tau \simeq (\pi/(8
\chi) \exp[4 \beta]$). We recall that in Kitaev's Toric code model on
the torus (the system with periodic boundary conditions)
the gap between the GS and  lowest excited energy levels is
equal to four: $\Delta =4$ [see the discussion after Eq. (\ref{thG})]. 
(Similarly, the gap for the model on an open surface (open boundary
conditions) is given by $\Delta =2$.) Thus, for Kitaev's Toric code model on a torus, in the limit of small
temperatures the inverse Botzmann factor $\exp[\beta \Delta]$ scales
in the same fashion as the equilibration time $\tau$. The important
feature of the scaling of the autocorrelation time in the Kitaev's
model is that, for large  systems, it is system size independent at all
temperatures. Similarly, in the long-time limit, 
\begin{eqnarray}
|t| \gg  \chi^{-1} \frac{1}{1 - \tanh 2 \beta},
\end{eqnarray}
the autocorrelation function is well approximated by
\begin{eqnarray}
\hspace*{-1cm}G_{X_{\mu};\mbox{\footnotesize \sf long-time}}(t) &=& \\
\hspace*{1cm} \sqrt{\frac{1+ \tanh 2 \beta}{2 \pi\chi |t| (1- \tanh 2
\beta)}}&& \hspace*{-0.5cm}e^{- \chi (1- \tanh 2 \beta) |t|} \nonumber.
\label{tauTT}
\end{eqnarray}
Identical relations hold for the autocorrelators $G_{Z_{\mu}}(t) \equiv
\langle Z_{\mu}(0) Z_{\mu}(t) \rangle$. Thus, as we increase the 
system size the error rate will always be finite at any $T>0$. In other
words, any topological quantum memory at finite (non-zero) temperatures
might not sustain self-correction indefinitely.

We now discuss Eq. (\ref{detectionT_cond}). It is readily  seen that
this condition is {\em violated} here. For instance, if we choose $V=
\sigma^{x}_{ij}$, and  $\hat{T}_\mu = Z_{1}$ with $(ij) \in C_{1}$  we
have, for any finite $\beta$, that 
\begin{eqnarray}
\rho^{1/2} [ \sigma^{x}_{ij}, Z_{1}] \rho^{1/2} \neq 0,
\end{eqnarray}
as $\{\sigma^{x}_{ij}, \sigma^{z}_{ij}\}=0$.  This is, of course, a
particular realization of the general result of Appendix \ref{nogo}.
{\em The Kitaev's Toric code model has a $T=0$ transition and thus cannot
satisfy the equilibrium thermal detection 
condition of Appendix \ref{nogo}.} It cannot be 
ruled out however that Kitaev's Toric code model may 
nevertheless satisfy quantum error detection
at finite temperature up to a small 
discrepancy nor that error detection may work
well for finite time intervals.
 
On the other hand, by choosing $V = \sigma^{x}_{ij}$, which has the
property that 
\begin{eqnarray}
\left ( 1+B_p \right ) \sigma^{x}_{ij} \left ( 1+B_p \right )=0,
\end{eqnarray}
and realizing that the GS projection operator is given by
\begin{eqnarray}
\hat{P}_{0} = \prod_{s} \left ( \frac{1+A_s}{2} \right ) \prod_{p} \left
( \frac{1+B_p}{2} \right ) ,
\end{eqnarray}
the $T=0$ detection condition, Eq. (\ref{detection_cond}), is
trivially  satisfied. Similarly, if one chooses $V = \sigma^{z}_{ij}$
since 
\begin{eqnarray}
\left ( 1+A_s \right ) \sigma^{z}_{ij} \left (
1+A_s \right )=0,
\end{eqnarray}
and, in general, for Kitaev's model Eq. (\ref{detection_cond}) is
satisfied. That the $T=0$ conditions  of Eq. (\ref{detection_cond}) are
satisfied also follows from  the fact that Eq. (\ref{def.}) implies
that Eq. (\ref{detection_cond}), and as we proved in Ref. \cite{NOLong}
Kitaev's model, satisfies the $T=0$ conditions of Eq. (\ref{def.}).

\subsection{Generalizations of Kitaev's Toric Code Model}
\label{Knew}

There is obviously no connection between the lack of thermodynamic
phase transition (as in Kitaev's model) and the existence of TQO as
defined in the Introduction. Indeed, we will now explore two models
that have TQO yet display finite-$T$ phase transitions as signaled by
non-analyticities in the free energy $F$. The first model we will
consider is a $D=3$ extension of Kitaev's model and the second a $D=3$
$\Z_k$ gauge theory. In both of these systems, no SSB occurs (and TQO
is indeed materialized). Nevertheless,  the system's free energy
displays singularities at finite temperatures.

\begin{figure}
\centerline{\includegraphics[width=0.7\columnwidth]{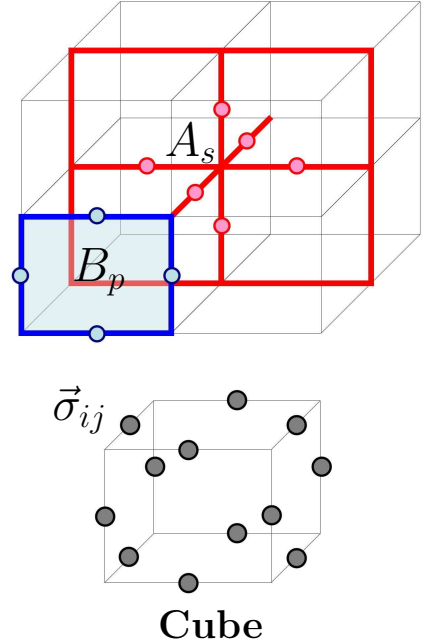}}
\caption{$D=3$ extension of Kitaev's Toric code Model. On each bond
$(ij)$, there is an $S=1/2$ degree of freedom indicated by a Pauli
matrix $\vec{\sigma}_{ij}$. Notice that elementary plaquettes are
planar while stars involve 6 neighbors in the $x$, $y$, and $z$ space
directions. A {\bf Cube} involves 6 plaquettes.}
\label{3dKitaev}
\end{figure}

The $D=3$ Kitaev's extended model (KE$3D$) in a cubic lattice has a
Hamiltonian formally written as  Eq. (\ref{kitaevmodel}) with star (or
vertex) operators $A_s$, each comprising the 6 nearest-neighbors to a
site, and planar plaquette operators $B_p$, each involving 4 spins (see
Fig. \ref{3dKitaev}). This Hamiltonian is basically a $D=3$ Ising gauge
theory ($-\sum_p B_p$) augmented by the sum of all local symmetry
generators  ($-\sum_s A_s$).  The constraints of Eq. (\ref{ABpc}) get
replaced by
\begin{eqnarray}
\prod_s A_s=1 \ , \ \prod_{p \in \mbox{\bf{Cube}}} B_{p} =1 ,
\end{eqnarray}
where {\bf Cube} includes the six plaquettes which form the cube. Since
no constraint couples the vertex and plaquette operators, the partition
functions is simply
\begin{eqnarray}
Z_{3D}=Z_{3D \ {\sf Ising ~gauge}} \times Z_{1D \ {\sf Ising}} .
\label{3Dpf}
\end{eqnarray}

Let us first show that KE$3D$ does show a thermodynamic phase
transition at $T>0$. From Eq. (\ref{3Dpf}) the free energy is given by
$F_{3D}(\beta)=F_{3D \ {\sf Ising ~gauge}}(\beta) + F_{1D \ {\sf
Ising}}(\beta)$, where  $F_{3D \ {\sf Ising ~gauge}}(\beta)=F_{3D \
{\sf Ising}}(\beta^*)$ with the dual $\beta^*$ satisfying  $\sinh 2
\beta \sinh 2 \beta^{*} = 1$. Clearly, the free energy of KE$3D$
displays a $T=0$ singularity coming from $F_{1D \ {\sf Ising}}(\beta)$
and a finite-$T$ singularity resulting from  $F_{3D \ {\sf Ising
~gauge}}(\beta)$ at $\beta_c=0.761423$.

We now show that the $D=3$ Ising Gauge theory displays (at least) rank
$n=8$ TQO. [See the definition of rank-$n$ TQO given in the
Introduction.]  Let us start by writing 8 GSs of the  $D=3$ Ising gauge
theory on a cubic lattice of size $L^3$ which is endowed  with periodic
boundary conditions
\begin{eqnarray}
| g_{+++} \rangle &=& {\cal{N}}  \sum_{c \in \{+++\}} 
|c \rangle, \nonumber \\ 
| g_{++-} \rangle &=& {\cal{N}}  \sum_{c \in \{++-\}} 
|c \rangle, \nonumber \\ 
&\vdots& \nonumber \\ 
| g_{---} \rangle &=&  {\cal{N}}  \sum_{c \in \{---\}} |c \rangle. 
\label{gangof8}
\end{eqnarray}

In Eqs. (\ref{gangof8}), the states $\{| c \rangle \}$ span all states
in the $\sigma_{ij}^{z}$ basis which  (i) have $B_{p} =1$ for each
plaquette $p$ [see the definition of $B_{p}$ in Eq. (\ref{AB_defn})] and
(ii) lie in a specific topological sector and ${\cal{N}}$  is a uniform
normalization constant. The eight topological sectors [$(+++), (++-),
\cdots, (---)$] are labeled by three Toric invariants ($\mu=1,2,3$)
\begin{eqnarray}
Z_{\mu} = \prod_{(ij) \in C_{\mu}} \sigma_{ij}^{z}.
\label{zaq}
\end{eqnarray}
Each of the $\bigotimes_{(ij)} \sigma^{z}_{ij}$ eigenstates $|c
\rangle$ is an eigenstate of the three Toric operators 
$\{Z_{\mu}\}_{\mu=1}^{3}$.  In Eq. (\ref{zaq}), each of the three
cycles $C_{1,2,3}$ is a path which circumscribes one {\it Toric} cycle
(e.g.  a cycle along each of the three cubic axes). Each of the states
in Eq. (\ref{gangof8})  transforms as a singlet under all of the cubic
lattice star operations 
\begin{eqnarray}
A_{s} = \prod_{i \in {\sf star}(s)} \sigma^{x}_{si},
\label{starlast}
\end{eqnarray}
with the product above performed over all 6 bonds $(si)$ which have the
vertex $s$ as one of their endpoints. This is so as the  star
operations of Eq. (\ref{starlast}) link states $|c \rangle$ within the
same topological sector. That is, if 
\begin{eqnarray}
A_{s} | c \rangle = |c ' \rangle
\end{eqnarray}
then, as $[Z_{\mu}, A_{s}]=0$ for all $\mu$ and $s$,
\begin{eqnarray}
Z_{\mu} | c \rangle = Z_{\mu} | c' \rangle.
\end{eqnarray}
Given the invariance of the 8 GSs of Eq. (\ref{gangof8}) [and thus of
any superpositions thereof] under all of the $d=0$ symmetries 
$\{A_{s}\}$, we may proceed to demonstrate TQO. To this end, let us 
decompose any quasi-local operator $V$ into the component invariant
under {\em all} of the $d=0$ GLSs of Eq. (\ref{starlast})  (labeled by
$V_{{\cal{A}};0}$) and component $V_{{\cal{A}}; \perp}$ which does not
transform as a singlet under all of these operations: $V= 
V_{{\cal{A}};0} + V_{{\cal{A}}; \perp}$. For any state $|g \rangle$
which lies in the 8-dimensional space spanned by the states of Eq.
(\ref{gangof8}),  and which transforms as a singlet under all $d=0$ 
GLSs ${\cal{G}}$, we have 
\begin{eqnarray}
\langle g | V_{{\cal{A}}; \perp} | g \rangle = 0.
\label{gvgp}
\end{eqnarray}
All that we need to consider are thus the local ($d=0$) symmetry
invariant components $V_{{\cal{A}};0}$ of the quasi-local  operator
$V$. The quasi-local operators invariant under all of the symmetries of
Eq. (\ref{starlast}) are built out of  product of a finite number of
operators $\{A_{s}\}$ and plaquette operators $\{B_{p}\}$ (see
Eq. (\ref{AB_defn})). We must now show that any such quasi-local
operator $V_{{\cal{A}};0}$  attains the same expectation value in  each
of the states of Eq. (\ref{gangof8}). To this end, we consider the
following three {\em connecting operators} 
\begin{eqnarray}
X_{\mu} = \prod_{(ij)  \perp \hat{e}_{\mu} }
\sigma^{x}_{ij}.
\label{TA}
\end{eqnarray}
These operators are the $D=3$ extension of the operators $X_{1,2}$ of
the $D=2$ $\Z_{2}$ gauge theory [Eq. (\ref{kit})].

In Eq. (\ref{TA}), the product is taken over all  bonds $(ij)$ which
lie in planes perpendicular  to the cubic direction $\hat{e}_{\mu}$. 
These operators commute with one another  and are symmetries of the
Ising gauge Hamiltonian: $[H, X_{\mu} ]=0$. Acting with a particular
$X_{\mu}$ on any state (e.g. any {\it vortex-less} $\sigma^{z}$
eigenbasis state $|c \rangle$) which is an eigenstate of all
$\{B_{p}\}$ operators with unit eigenvalue leads to states which are
eigenvectors of $\{B_{p}\}$ with unit eigenvalue. More generally, for
any operators $A_{s}$ and $B_{p}$, we have 
\begin{eqnarray}
[A_{s}, X_{\mu} ] = [B_{p}, X_{\mu} ]=0.
\label{ABT}
\end{eqnarray}
As a consequence of Eq. (\ref{ABT}) and the fact that all quasi-local
operators $V_{{\cal{A}};0}$  are multinomials in $\{A_{s}, B_{p}\}$, we
have that 
\begin{eqnarray}
[V_{{\cal{A}};0}, X_{\mu}]= 0.
\label{VAT}
\end{eqnarray} 
Moreover, these operators satisfy the following algebra with respect to
the Toric symmetries $\{Z_{\mu}\}$:
\begin{eqnarray}
\{X_{\mu}, Z_{\mu} \} = 0 \ , \ [X_{\mu}, Z_{\nu}] = 0 \  , \ \mu \neq
\nu .
\label{tz}
\end{eqnarray}
As a consequence of Eq. (\ref{tz}), we see that  the 8 states of
Eq. (\ref{gangof8}) are related to one another by these operators. For
instance, 
\begin{eqnarray}
|g_{++-} \rangle &=& X_{3} | g_{+++} \rangle, \nonumber \\ 
|g_{-+-} \rangle &=& X_{1} X_{3} |g_{+++} \rangle,
\end{eqnarray}
etc.. Therefore,  the 8 $d=1$ GLS operators $\prod_{\mu}
X_{\mu}^{n_{\mu}}$ with $n_{\mu} = 0,1$ form a $d=2$ group
${\cal{G}}$ (of a  $\Z_{2} \times \Z_2 \times \Z_{2}$ character). These
operators suffice to link all of the states of Eq. (\ref{gangof8})
with  one another. By unitarity, these generators also link any set of
8 orthogonal states in the space spanned by Eq. (\ref{gangof8}). 

By Eqs. (\ref{gvgp}, \ref{VAT}), the expectation value of any
quasi-local operator $V$ is the same in all GSs spanned by the $n=8$
GSs of Eq. (\ref{gangof8}). Thus, the $D=3$ Ising gauge theory exhibits
(at least) rank-$n=8$ TQO. 

We note, in passing, that considerations similar to those above may be
enacted for general $\Z_{k}$ gauge theories ($k =2,3,4, \cdots$) on a
hypercubic  $D$-dimensional lattice. This theory is defined by 
\begin{eqnarray}
H_{\Z_k} = - \frac{1}{2} \sum_{p} (U_{ij} U_{jk} U_{kl} U_{li} + {\rm
h.c.})
\end{eqnarray}
where, on every link $(ij)$, a parallel transporter
\begin{eqnarray}
U_{ij} &=& e^{i \theta_{ij}}, \nonumber \\ 
\theta_{ij} &=& 2 \pi n_{ij}/k
\label{uijkd}
\end{eqnarray}
with $n_{ij}$ an arbitrary integer. The elements $U_{ij}$ of Eq.
(\ref{uijkd}) satisfy a $\Z_{k}$ algebra. Here, instead of Eq.
(\ref{TA}), we set
\begin{eqnarray}
\hat{T}_{\mu} =  \prod_{(ij)  \perp \hat{e}_{\mu} } e^{i \frac{2
\pi}{k}  L^{z}_{ij}}.
\label{tad}
\end{eqnarray}
In Eq. (\ref{tad}), $L^{z}_{ij}$ is the generator of rotation of the
$\Z_{k}$ variable on bond $(ij)$.  Replicating the proof given above,
we find that the $\Z_{k}$ gauge theory on the $D$-dimensional lattice
exhibits (at least) rank-$n=k^{D}$ TQO. 

It remains to prove that at any finite $T>0$ no topological symmetry
operator may acquire a non-vanishing expectation value. For example, 
\begin{eqnarray}
\langle Z_{C_{\mu}} \rangle =  \langle \prod_{(ij) \in C_{\mu}}
\sigma^{z}_{ij} \rangle =0,
\label{zzzz}
\end{eqnarray}
for $\{C_{\mu}\}$ loops around the Toric cycles.  [Equation
(\ref{zzzz}) is a particular realization  of Eq. (\ref{central}).] Due
to the decoupling of the plaquette $(\{B_{p}\}$) and vertex
($\{A_{s}\}$)  operators, the expectation value $\langle Z_{C_{1}}
Z_{C_{2}} \rangle$ is given by its value for a classical $D=3$ Ising
gauge theory. However, as seen by large and small coupling expansions
\cite{kogut}, the correlator $\langle Z_{C_{1}} Z_{C_{2}} \rangle =
\langle  \prod_{(ij) \in C_{1}} \sigma^{z}_{ij}  \prod_{(ij) \in C_{2}}
\sigma^{z}_{ij}  \rangle$  vanishes in the $D=3$ Ising gauge theory as
the bounding contours $C_{1,2}$  are taken to be infinite.  [For finite
size systems, $\langle Z_{C_{\mu}} \rangle$  vanishes as no SSB is
possible.]  In (i) the {\it confined} phase  [$\beta < \beta_{c2}$]
this correlator vanishes exponentially in the area of the minimal
surface $R$  bounded by the Toric cycles $C_{1,2}$ while in (ii)  the
{\it deconfined} phase [$\beta> \beta_{c2}$], the pair correlator 
$\langle Z_{P_{1}} Z_{P_{2}} \rangle$ vanishes exponentially in the
total length of the contours $C_{1}$ and $C_{2}$. In the limit of far
separated contours $C_{1}$ and $C_{2}$, both (i) and (ii) reaffirm Eq.
(\ref{zzzz}). Equation (\ref{zzzz})  is suggestive of a finite
autocorrelation time at all positive temperatures.

\subsection{Kitaev's Honeycomb Model}
\label{kith}

\begin{figure}[h]
\includegraphics[width=3.4in]{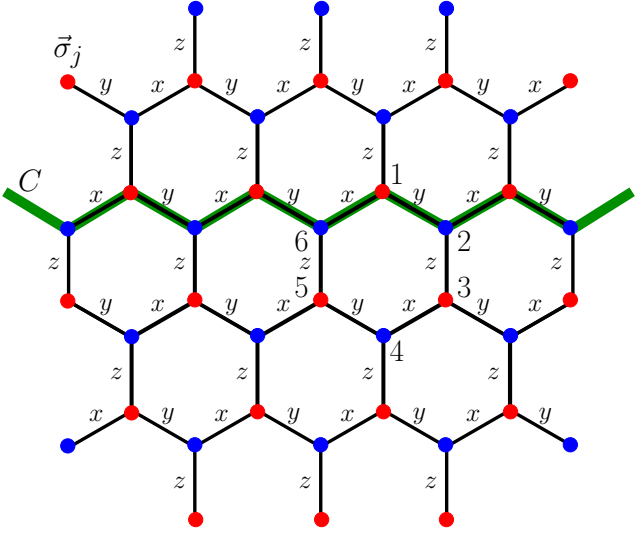}
\caption{Kitaev's model on a honeycomb lattice and three  types of
bonds. On each vertex there is an $S=1/2$ degree of freedom indicated
by a Pauli matrix $\vec{\sigma}_{j}$. $C$ represents an arbitrary
contour drawn on the lattice.}
\label{figk}
\end{figure}

Kitaev's model on the honeycomb lattice \cite{Kitaev2006} is  defined
by the following $S=1/2$ Hamiltonian (Fig. \ref{figk})
\begin{eqnarray}
H_{K_h}\!\!=\!\!-J_x \!\!\!\!\sum_{x{\sf -bonds}}
\!\!\!\sigma^x_{j}\sigma^x_{k} -J_y\!\!\!\!\sum_{y{\sf -bonds}}
\!\!\!\sigma^y_{j}\sigma^y_{k} -J_z\!\!\!\!\sum_{z{\sf -bonds}}\!\!\!
\sigma^z_{j}\sigma^z_{k}.
\label{H}
\end{eqnarray}

Here, we find that Elitzur's theorem mandates that {\em all}
non-vanishing expectation values must be of the form \cite{CN}
\begin{eqnarray}
\hat{O}_{C} = \prod_{j \in C} \sigma_{j}^{\gamma},
\label{oc}
\end{eqnarray}
with $C$ any contour (set of contours) drawn on the lattice and $
\gamma$ is the bond direction which is orthogonal to the path $C$. Let
us consider embedding this system on a torus with $g$ handles (a torus
of genus $g$). For closed contours $C$ which do not span  an entire
Toric cycle, $\hat{O}_{C}$ is not independent of the code's stabilizer.
In fact, there is a rather simple connection between  these operators
and the local anyon charge which measures the  number of local high
energy defects. For a closed contour $C$ which is an elementary closed
hexagonal loop, we have $\hat{O}_{h}$ which is the anyon charge
associated with a given hexagon $h$. Rather specifically, 
\begin{eqnarray}
\hat{O}_{h} = \sigma^{z}_{1} \sigma^{x}_{2} \sigma^{y}_{3}
\sigma^{z}_{4} \sigma^{x}_{5} \sigma^{y}_{6}
\label{hexagonaa}
\end{eqnarray}
with $1-6$ labeling the vertices  of any given hexagon of Fig.
\ref{figk}. The spin polarization directions $\mu$ for all spins
$\sigma^{\mu}_{i}$ in the product of Eq. (\ref{hexagonaa}) have been
chosen to correspond to the single bond direction $\mu$ ($x,y,$ or $z$)
that is attached to the site $i$ and does not lie on the hexagonal
path.

As shown by Kitaev, within the GS sector, $\langle \hat{O}_{h}\rangle =
1$ for all $h$. Extending the arguments of the strong and weak coupling
expansions of gauge theories \cite{kogut}, we  now find that the
correlator between any two contours $\langle \hat{O}_{C} \hat{O}_{C'}
\rangle$  scales as (i) $e^{-c_{1} A}$ at high temperatures with $A$
the area bounded by $C$ and $C'$ and $c_{1}$ a positive constant and,
if an ordered phase exists, scales  at low temperatures as (ii)
$e^{-c_{2} (|C| + |C'|)}$ with  $|C| + |C'|$ the total perimeter of the
two loops and $c_{2}$ another positive constant. Thus, for cycles $C$
which span the entire lattice we find as before that no $d=1$ loops of
the form of Eq. (\ref{oc})  can attain a finite expectation value.  
      
Let us now consider closed contours $C_{1,2}$ which span the  same
Toric cycle.  For a system with periodic boundary conditions, at $T=0$,
we now have for closed contours $C_{1,2}$, 
\begin{eqnarray}
\langle  \prod_{i \in C_{1}} \sigma^{z}_{i}  \prod_{j \in C_{2}}
\sigma_{j}^{z} \rangle = \langle \prod_{h \in {\sf D}}  \hat{O}_{h}
\rangle  =1.
\label{c1c2h}
\end{eqnarray}
Here, ${\sf D}$ is the domain bounded by two contours $C_{1,2}$ which 
go around an entire cycle of the torus [see e.g. the thick solid line
in  Fig. \ref{figk}]. Taking the separation between the two contours
$C_{1}$ and $C_{2}$ to be very large,  we have that, at $T=0$, 
\begin{eqnarray}
\langle \prod_{i \in C_{1}} \sigma^{z}_{i} \prod_{j \in C_{2}}
\sigma_{j}^{z} \rangle  \to \langle  \prod_{i \in C_{1}} \sigma^{z}_{i}
\rangle \langle \prod_{j \in C_{2}} \sigma_{j}^{z}  \rangle.
\end{eqnarray}
This implies that, at $T=0$, 
\begin{eqnarray}
\langle \prod_{i \in C_{1}} \sigma^{z}_{i} \rangle = \pm 1.
\label{zeroT}
\end{eqnarray}
At $T>0$, no SSB may occur (also in the thermodynamic limit, see Eq.
(\ref{perimeter})) and all quantities of the form Eq. (\ref{zeroT})
must be zero.  

We now write down a formal finite temperature solution to the partition
function and comment on low- and high-$T$ series expansions. Using
fermionization \cite{CN}, Kitaev's Honeycomb model can be cast as a
model for fermions $\{d_{r}\}$ on a square lattice with a site-dependent 
chemical potential $\eta_{r}= \pm 1$.
\begin{eqnarray}
H_{K_h} &=&  J_x\sum_{r}\left(d_r^\dag + d_r^{}\right) \left(
d_{r+\hat{e}_x}^\dag - d_{r+\hat{e}_x}^{} \right) \nonumber \\
&+&J_y\sum_{r}\left(d_r^\dag + d_r^{}\right) \left( d_{r+\hat{e}_y}^\dag -
d_{r+\hat{e}_y}^{} \right) \nonumber \\
&+&J_z\sum_r \eta_r (2d^\dag_r d^{}_r-1) \nonumber \\ 
&\equiv& d^{\dagger} M_{\eta} d + d^{\dagger} W_{\eta} 
\tilde{d}^{\dagger} + \tilde{d} W_{\eta}^{\dagger} d.
\label{EQ-fermion-model}
\end{eqnarray}
The fermionization of \cite{CN} has been recently invoked  to attain
very interesting results in extensions of Kitaev's Honeycomb model
\cite{Yu}. In Eq. (\ref{EQ-fermion-model}), $\{\vec{r}\}$ denote the
centers of the  vertical bonds. The unit vector $\hat{e}_y$ connects
two $z$-bonds and crosses a $y$-bond, see  Fig. \ref{figk}. A similar
definition holds for $\hat{e}_x$ \cite{CN}. The form of Eq.
(\ref{EQ-fermion-model}) is very similar to the Fermi representation of
the $D=2$ Ising model. In the last line of Eq.
(\ref{EQ-fermion-model}),
\begin{eqnarray}
d^{\dagger} = (\{d_r^{\dagger}\}),~~ \tilde{d} = (\{d_{r}\}).
\end{eqnarray}
With $\Lambda =  (d^{\dagger}, \tilde{d})$ \cite{Pan} and
\begin{eqnarray}
N_{\eta} = \left( \begin{array}{cc}
    M_{\eta} & 2 W_{\eta} \\
   2W_{\eta}^\dagger & -M_{\eta} 
\end{array} \right),
\end{eqnarray}
the partition function for the Fermi bilinear of Eq.
(\ref{EQ-fermion-model}) is easy to write down
\begin{eqnarray}
Z_{K_h} = \sum_{\{\eta_{r}\}} e^{-\frac{\beta}{2} \tr [M_{\eta}]} 
|\det (e^{\beta N_{\eta}} + 1)|^{1/2}.
\label{Ztr}
\end{eqnarray}
Within the GS sector (the one with no anyons), the system is
translationally invariant and  its spectrum \cite{Kitaev2006} is that
of a $p$-wave type BCS pairing problem \cite{CN,Yu}. 

Due to the sum over all of the $2^{N_{s}/2}$ Ising  configurations
$\{\eta_{r}\}$, the complete partition function of Eq. (\ref{Ztr}) is
non-trivial.   The existence of a transition as temperature is varied
is not as immediate as in Kitaev's Toric code model. Translational
invariance appears only for uniform $\eta_{r} =1$ or $-1$  for all $r$
and by a simple unitary  transformation to a system in which $\eta_{r}$
is constant on entire horizontal lines \cite{CN}. Nevertheless, bounds
on correlators are easily established: any fermionic correlator
computed with the  full partition function of Eq. (\ref{Ztr}) is
bounded from above by its value when computed within the sector
$\{\eta_{r}\}$ which maximizes its value. The finite $T$ fixed sector
correlators for the quadratic Hamiltonian of Eq.
(\ref{EQ-fermion-model}) can be computed with the aid of Wick's
theorem. If the GS sector (that with $\eta_{r}=1$ everywhere) is gapped
then at all temperatures, the correlation functions exhibit
exponentially decaying correlations. 

We may expand Eq. (\ref{Ztr}) in low- and high-$T$ series. A  high-$T$
series may be derived by  an expansion of $|\det (e^{\beta N_{\eta}} +
1)|$  in powers of $\beta$. At low temperatures,  the GS terms
correspond to the phase $\eta_{r}=1$ for all $r$ which just reproduces
the $p$-wave type BCS result of \cite{Kitaev2006,CN}. At finite
temperatures, we allow for $\eta_{r} \neq 1$. The lowest energy terms
correspond to a few vortex pairs ($\eta_{r} = -1$) which are tightly
bound. We may diagonalize the Hamiltonian to determine the spectrum for
these vortex patterns and find the low-$T$ corrections to the GS
results. In such a manner, we may write down a low-$T$ series expansion
for the correlation function between fermion pairs (which corresponds
to string correlators in terms of the spin variables of Kitaev's
model). The explicit high- and low-$T$ series expansions will be
provided elsewhere. For now, let us note that within  the gapped phase,
the system exhibits a finite correlation length for these string
correlators \cite{CN}. It is clear that if a low-$T$ series expansion
about the GSs is possible, then at sufficiently low temperatures,
corrections to the $T=0$ result for the string correlators can be made
arbitrarily small. In particular, when such an expansion is carried 
out about the GSs of the $T=0$ gapped phase,  the string type
correlators of \cite{BN} still  display a finite correlation length.
[As we discussed above, that these string correlators must indeed 
display a finite correlation length also follows from  the form of the
fermionic correlator in the $\{\eta_{r}\}$ sector which maximizes its
value; this value of the correlator provides  an upper bound on the
correlation function of the complete (unpinned $\{\eta_{r}\}$) system].
The correlation function  on the righthand side of Eq. (\ref{zeroT}) is
related to a limiting form of such string correlators in which the
string (defined  here by the contour $C_{1}$) wraps around an entire
Toric cycle.

In what follows, we comment on the prospect of finite-$T$ phase 
transitions in Kitaev's Honeycomb model as it follows from our
expansions. As we show in the Appendix, in the absence of a phase
transition, the topological quantum error detection condition of Eq.
(\ref{detectionT_cond}) [and its equivalent Eq.
(\ref{detectionT_cond+})], cannot be satisfied. If no transitions
exist, then hardware fault-tolerance at finite temperatures may be
impossible to achieve. The $T=0$ gapless phase - the phase which is not
robust - seems to be amenable to a phase transition as the temperature
of the system is increased.  In the gapless phase,  a {\em transition}
from a system with oscillatory algebraic  correlation between fermions
at $T=0$ to one with exponential correlations at high temperatures is
suggested by the low- and high-$T$ expansions. For states deep inside
the gapless region \cite{Kitaev2006,CN} a low-$T$ expansion (that
includes sectors with a non-vanishing number of anyons) suggests that
low-energy states may still carry a vanishing gap and lead to a $T>0$
phase with algebraic correlations. Thus, a finite-$T$ transition
between a high-$T$ phase with exponentially decaying correlations
between fermions to a low-$T$ {\it stripe type} phase with oscillatory
algebraic correlations may exist. 

In the absence of applied external fields which introduce a gap, the
gapless sector of the theory is, however, unstable to perturbations. Of
greater promise -- insofar as stability to local perturbations --  is
the gapfull region of the Kitaev model. Regretfully, no transition of
the type found for the gapless phase is evident for the gapfull sector
of the Honeycomb model.  Thus, albeit satisfying the $T=0$ condition
of  Eq. (\ref{detection_cond}), in the absence of a finite-$T$ phase
transition, the condition of Eq. (\ref{detectionT_cond}) cannot be
satisfied at any finite temperature. 

Our results on the low- and high-$T$ expansions, which suggest that the
gapfull sector of the Kitaev model does not exhibit a finite-$T$ phase
transition, can be further fortified and linked to  expansions carried
by earlier works. These expansions elucidate how the gapfull phase will
not break ergodicity and cannot achieve hardware fault-tolerance. The
correspondence to our earlier results concerning Kitaev's Toric code
model is made clear by a simple mapping. In the parameter region $J_{z}
\gg J_{x,y}$, a perturbative expansion gives the following effective
Hamiltonian \cite{Kitaev2006}
\begin{eqnarray}
H_{\sf eff} \simeq - \frac{J_{x}^{2} J_{y}^{2}}{16 J_{z}^{3}} (\sum_{s}
A_{s} + \sum_{p} B_{p}).
\end{eqnarray}
In other words, we recover Kitaev's Toric code model of Eq. 
(\ref{kitaevmodel}). As a result, we have as before 
that the logical operators ($X_{1,2}$ and $Z_{1,2}$) have a finite
(system-size independent) autocorrelation time $\tau$ [Eq. (\ref{gx5})]. 

\section{Thermally stable orders in the case of high-dimensional
broken symmetries - A case of simple non-TQO}

Some systems with non-local symmetries $\hat{T}_{\mu}$ exhibit a 
robustness to thermal fluctuations. These systems are natural
contenders for satisfying the conditions of Eq.
(\ref{detectionT_cond}). In all of the cases that we examined,
however,  thermal fragility still appears.

Let us start with the simple $D=2$ orbital compass model. Its
Hamiltonian is given by
\begin{eqnarray}
H = - \sum_{j} [J_{x}\sigma^{x}_{j} \sigma^{x}_{j+ \hat{e}_{x}} + J_{z}
\sigma^{z}_{j} \sigma^{z}_{j+\hat{e}_{z}}],
\end{eqnarray}
which  emulates the direction dependent interactions resulting from the
anisotropy of electronic orbitals. In this model, exchange interactions
involving  the $x$ component of the spin occur only along the spatial
$x$-direction of the lattice. Similar spatial direction dependent spin
exchange interactions appear for the $z$ components of the spin. Apart
from a global reflection which only appears for the isotropic  point
$J_{x} = J_{z}$ (and which may be broken at low $T$),  the anisotropic
orbital compass model has the following  symmetry operators
$\hat{O}_{\mu} =  \prod_{j \in C_{\mu}} i\sigma_{j}^{\mu}$ for $\mu =
x,z$.  $C_{\mu}$ denotes any line orthogonal to the $\hat{e}_{\mu}$
axis. As $\hat{O}_{\mu}$ involves ${\cal{O}}(L^{1})$ sites, they
constitute $d=1$ symmetries. The spectrum of the $D=2$ orbital compass
model is gapless \cite{mila} in the thermodynamic limit due to the
existence of the symmetries that it possesses \cite{NOLong,NF,BN}. What
is important for relaxation processes  (much as was evident in the
analysis of Kitaev's model) is not at all the energy spectrum but
rather the free energy. It was found that the large $S$ renditions of
the isotropic $D=2,3$-dimensional orbital compass model, although
highly degenerate in the low-energy spectra in the thermodynamic limit,
exhibits free energy barriers. An entropic stabilization sets in which
at sufficiently low yet finite $0<T<T^{*}$ leads to a sharp phase
transition with a divergent relaxation time \cite{NBC}. It is
noteworthy, however, that at the isotropic point, the system is not
topologically ordered as a global inversion symmetry 
\begin{eqnarray}
\hat{O}_{\sf Reflection} = \prod_{j} e^{i \frac{\pi \sqrt{2}}{4}
(\sigma_{j}^{x} + \sigma_{j}^{z})}
\label{reflbad}
\end{eqnarray}
can be broken. It is this very symmetry which is broken below $T^{*}$. 
Here, the local nematic order parameter operator 
\begin{eqnarray}
N_{j} =  \Big[\sigma^{x}_{j} \sigma^{x}_{j+ \hat{e}_{x}} -
\sigma^{z}_{j} \sigma^{z}_{j+\hat{e}_{z}} \Big]
\label{N}
\end{eqnarray}
attains a finite expectation value at low $T$ \cite{NOLong,NBC,NF,BN}
\begin{eqnarray}
\langle N_{j} \rangle_{\alpha} \neq \langle N_{j} \rangle_{\beta}, ~~~~
T<T^{*} .
\end{eqnarray}
Thus, measurements of this quasi-local order parameter within the GS
sector lead to different answers for different contending low-energy
Gibbs states (labeled by $\alpha$ and $\beta$). Thus, $N_{j}$ is not a
contending logical operator - it does not satisfy Eq. (\ref{VTVT}). For
all non-local operators $\hat{O}_{\mu}$ we have that  at any finite
$T$, $\langle \hat{O}_{\mu} \rangle =0$  on any lattice (whether it is
finite and no SSB is possible or on the infinite size lattice where the
perimeter law scaling implies a vanishing  $\langle \hat{O}_{\mu}
\rangle =0$).

Similar arguments apply for the $D=3$ orbital compass model, meaning
that the $d=2$  symmetries (as well as a three-dimensional reflection
symmetry) will render the TQO non-existent at finite temperatures,
although the relaxation time is obviously divergent. We recently became
aware of a paper by Bacon \cite{bacon} where he provides evidence (by
using a mean-field approach) of a self-correcting system in a $D=3$ 
relative of the orbital compass model. The reason why this system is
claimed to be a high-temperature self-correcting quantum memory (i.e.
one without additional quantum error correction) is  because, at the
mean-field level, low-energy excitation energies may scale with the
perimeter of the defected domain. As in the $D=3$ orbital compass
model, Bacon's model display $d=2$ (discrete) $d$-GLSs. The problem we
see with this argument is that these symmetries can be broken giving
rise to an ordered phase (i.e. one with a Landau order parameter).
Then, condition (\ref{def.}) is going to be violated and no TQO will be
present. As in other cases in which symmetries  can be broken, here Eq.
(\ref{VTVT}) - a version of which is related to the weaker version [Eq.
(\ref{VTV})] of the finite-$T$ error detection condition of Eq.
(\ref{detectionT_cond}) - will be violated.

\section{Can thermal fragility be defeated?}

It is clear that a physical system that is in an ordered thermodynamic
phase, i.e. having a non-vanishing Landau order parameter (i.e.
breaking a global symmetry), displays some sort of robustness against
{\it local} errors due to the {\it collective} nature of the resulting
order. This property is used to build robust {\it classical memories}.
We have seen that topological quantum memories seem to be thermally
fragile, therefore, it seems plausible that by building systems that
display an order parameter we could defeat it and obtain robust {\it
quantum memories}.  In general, there seems to be a catch regarding
stable quantum topological memories. According to Eq.
(\ref{detectionT_cond}), in order to ensure error detection of local
perturbations at finite temperatures, the system must display at least
one finite temperature phase transition (see Appendix \ref{nogo}). We
generally find that $\langle \hat{T}_{\mu} \rangle =0$. As a matter of
principle, even if we allow for operators $\{\hat{R}_{a}\}$ of a
topological character (that is, non-local operators) which are not
independent of the code's stabilizer,  unless a low-$T$ series
expansion about the $T=0$ state is void, we will always get a
perimeter-law type scaling for the quantum topological observables, see
e.g. Eq. (\ref{perimeter}). This expansion for local Hamiltonians [see
e.g. Ref. \cite{kogut}] always gives us the result that the topological
observables vanish for large loops (for which local errors  can be
avoided at $T=0$).  The expansion about $T=0$ can be void if there is a
$T=0$ transition on an infinite size system. Such an occurrence 
matters worse once again as  it implies that the robustness at $T=0$
does not obviously imply a robustness at any finite temperature (even
at temperatures $T=0^{+}$). On any finite size system, there is no SSB
and as before $\langle \hat{T}_{\mu} \rangle =0$ identically. We
illustrated that the same also occurs in the thermodynamic limit of
these systems. Ergodicity cannot be broken insofar as the logical
operators $\{\hat{T}_{\mu}\}$ can detect it.

We do not see an obvious way to avoid this conundrum both for the
viable logical operators $\{\hat{T}_{\mu}\}$ which are independent of
the code's stabilizer and  more generally also for general non-local
operators $\{\hat{R}_{a}\}$ that can encode topological operations. It
should be emphasized that all of our results pertain to rigorous
bounds. Although finite (and size-independent) autocorrelation times
seem to be mandated,  if situations exist in which $\Delta \gg k_{B} T$
may be realized then the autocorrelation times can be made large (see
e.g. the explicit form of Eq. (\ref{tauT})).  Based on heuristics and
static arguments, a recent review provides estimates on these ratios in
some candidate systems \cite{ds}.

\section{Summary}

In the present article we expanded on the concept of thermal fragility
and applied it to several case examples including Kitaev's Honeycomb
model and extensions to higher spatial dimensions. Our main results are

(i) We introduced and discussed quantum error detection at  finite
temperature as it applies to topologically ordered systems [Eq.
(\ref{detectionT_cond})].

(ii) We find that a system cannot satisfy the finite temperature error
detection criteria of Eq. (\ref{detectionT_cond}) unless it displays a
finite temperature phase transition. It is important to emphasize that
the existence of a finite temperature transition (or several finite
temperature transitions) is a necessary but not  sufficient condition
for  Eq. (\ref{detectionT_cond}) to hold.

(iii) In general TQO systems, whether they display finite-$T$ phase
transitions or not, and whether they have a finite size or are systems
taken in the thermodynamic limit, all non-trivial topological operators
may have a vanishing expectation value at all non-zero temperatures:
$\langle \hat{T}_{\mu} \rangle =0$. This suggests that the  code might
{\it self-correct} only up to finite cutoff  times that are governed by
temperature dependent effects. Depending on the relative size of the
gaps to the temperature, this autocorrelation time can practically be
made very large. This may go hand in hand with (system size
independent) finite bounds on the autocorrelation  times $\tau$ as
adduced from the values of  $|\langle \hat{T}_{\mu}(0) \hat{T}_{\mu}(t)
\rangle|$. For times $|t| \gg \tau$, $|\langle \hat{T}_{\mu}(0)
\hat{T}_{\mu}(t) \rangle|$ tends to zero [Eqs. (\ref{badmemory},
\ref{gx5}, \ref{tauT}, \ref{tauTT})].  Information stored in the
correlators of $\{\hat{T}_{\mu}\}$ may be lost. Insofar as the
non-local {\it topological} operators $\{\hat{T}_{\mu}\}$ can detect,
the system becomes ergodic at large time. Although of, apparently, less
pertinence to ideal quantum memories, the same result is found for 
general operators $\{\hat{R}_{a}\}$ which  encode (non-local)
topological operations  on infinite-size systems but are not
necessarily independent of  the code's stabilizer, Eq.
(\ref{perimeter}), or correspond to symmetries. 

(iv) Kitaev's Toric code model can be solved exactly. At all non-zero
temperatures, the autocorrelation function exhibits a finite
equilibration beyond which ergodicity sets in and all information is
lost. The explicit forms for the autocorrelation functions
of Kitaev's Toric code model are given in Eqs. (\ref{gx5}, \ref{tauT},
\ref{tauTT}).

(v) Kitaev's Honeycomb model can formally be solved  at finite
temperatures. The main utility of this formal solution is that it
enables low- and high-temperature series expansions. These  expansions
suggest that this model may  not be stable to thermal fluctuations (in
the sense of Eq. (\ref{detectionT_cond})). 

(vi) We showed that some candidate systems which  are stable to thermal
fluctuations often display local orders. These local orders render the
system unstable to errors from local perturbations and the {\it
topological protection} is lost.

A vexing question is how to quantify topological quantum error
conditions at finite temperatures. The connection between  rigorous
bounds on quantum error detection and  the more physical (although
rigorous)   results presented in our work  is far from clear. In our
opinion, this question needs to be quantitatively addressed. Given the
physical results that we derived in our work, it is unclear to us if
there is no way to  overcome thermal fluctuations and still maintain
quantum self-correction that is immune to local perturbations. We hope
that our work will stimulate further studies which will address many
related questions: Can $d=2$ GLSs lead to robust quantum memories? What
is the relation between a finite lifetime $\tau$ and the vanishing
expectation value of the logical operators $\hat{T}_{\mu}$? What
happens when a certain amount of disorder is present and thus, in
general, $d$-GLSs are not (exactly) present? In that case it is not
true that for a finite system the expectation value of the logical
operators vanishes since, for example for Kitaev's Toric code model
$[H_K+V, Z_\mu]\neq 0$ where $V$ represents the disorder. We conjecture
that, nevertheless, in the thermodynamic limit those expectation values
vanish,  for example $\langle Z_\mu \rangle \rightarrow 0$, and more
importantly the autocorrelation times can generally remain finite as
long as $V$ is an arbitrary  quasi-local operator (and the amount of
disorder is not big). Indeed, in our earlier work \cite{NOLong}, we
showed that system's thermodynamic behavior and response  functions may
remain adiabatic even when  their $d$-GLSs may be broken.

The main message is that we need to be careful when designing a
topological quantum memory. We are forced to consider finite
temperature effects since a real system is subjected to temperature
effects. The common lore that a finite gap can indefinitely protect 
states (and thus quantum information) is not always correct.  The
limiting size of the autocorrelation time can often be system
independent. If very low temperatures may indeed be achieved relative
to the gap \cite{ds} then the autocorrelation time cutoff (e.g. Eq.
(\ref{tauT})) can be made large (albeit still finite). It may also be
that systems with  a well crafted {\em environment} can be constructed;
such  environments will disallow rapid thermal equilibration and  thus
keep the autocorrelation times high. Possible realizations of such
environments can include a frequency mismatch between the natural
resonant frequencies of the TQO system and  the environment in which it
is embedded \cite{Loss}.

In some of the Appendices that follow, we elaborate on points related
to the main results reported here. In Appendix \ref{nogo}, we prove
that in the absence of a phase transition (a singularity in the free
energy), the finite temperature error detection condition of Eq.
(\ref{detectionT_cond}) might not hold. In Appendix \ref{LRS}, we
discuss a trivial consequence of the Lieb-Robinson bounds to {\it
topological} quantities. In Appendix \ref{anyp}, we sketch how rigorous
bounds for the excitation of anyons may be  derived in some disorder
free systems in thermal equilibrium.  The derivation outlined here  is
intimately linked to Kitaev's Honeycomb model of  Section \ref{kith}.
However, the result is far more general. In Appendix \ref{equilta}, we
discuss thermal equilibration time (the time for the system to become
ergodic) in cases when the  system displays or is close to a critical
point and general scaling laws become apparent.

{\em Note added in proof.}
After completing the current work and much after \cite{NOLong},  we
became aware of the interesting independent work  of Ref. \cite{alicki}
which raised concerns about  the reliable storage of quantum
information at  finite temperature. Reference \cite{alicki} invoked 
the KMS equations in the study of Kitaev's Toric code  model to
independently also arrive at and fortify one of the conclusions of Ref.
\cite{NOLong} (that of  the last line of Eq. (\ref{expectpart})) from a
different approach.

\acknowledgments 

We thank the Perimeter Institute and the IQC for their hospitality.  Manny
Knill and John Preskill are acknowledged for helpful discussions. This
work was sponsored, in part, by the CMI at Washington University, St.
Louis (ZN).

\appendix 

\section{Topological quantum error detection at finite temperatures}
\label{nogo}

We show that any system that does not exhibit a finite temperature
($T>0$) phase transition (regardless of whether it exhibits  TQO or
not) necessarily violates the quantum error detection condition of 
Eq. (\ref{detectionT_cond}) at {\em all} temperatures $T>0$. In other
words, for all $T>0$
\begin{equation}
W=[\rho^{1/2} V \rho^{1/2}, \hat{T}_\mu] \neq 0 .
\label{detectionT_cond+}
\end{equation} 
Here, we recall that $\rho ={\cal Z}^{-1} \exp[-H/(k_{B}T)]$ with
${\cal Z} = \tr [\exp[-\beta H]] = \exp[-\beta F]$ with $F$ the free
energy. In the absence of a phase transition, the free energy may be
expanded about its high temperature (small $\beta$) limit
\begin{eqnarray}
F = \sum_{k=0}^{\infty} F_{k} \beta^{k}. 
\label{s2}
\end{eqnarray}
All terms in the series of Eq. (\ref{s2}) are $c$-numbers. If the
system displays a phase transition the expansion of Eq. (\ref{s2}) has
a finite radius of convergence ($\beta_{c}$). If the system displays
no phase transition, the series of Eq. (\ref{s2})
is everywhere convergent.

The proof of the assertion of Eq. (\ref{detectionT_cond+}) for the case
a divergent  $\beta_{c}$ (a system with no finite temperature
transitions)  is trivial. For any finite system at temperatures $T>0$,
the density  matrix $\rho$ can be expanded to all orders in
$\beta=1/(k_BT)$. We have that ($\rho =\exp[-\beta(H-F)]$)
\begin{eqnarray}
W \!\!\!&=& \!\! [V,\hat{T}_{\mu}] - \frac{1}{2}  \beta
[\{(H-F_{0}),V\},\hat{T}_{\mu}]  + \frac{\beta^{2}}{2}
[\{F_{1},V\},\hat{T}_\mu] \nonumber \\
+&& \!\!\!\!\!\!\!\!\!\! \frac{\beta^{2}}{8}
[\{(H-F_{0}),\{(H-F_{0}),V\}\},\hat{T}_{\mu}]  +{\cal{O}}(\beta^{3}) 
.
\label{quantum_expander}
\end{eqnarray}
At any fixed chosen $\beta$ for which $F(\beta)$ is non-singular, the
expansion of $\exp[-\beta(H-F)]$ in powers of $\beta$,  with $F(\beta)$
held at its fixed value at the chosen $\beta$, has an infinite radius
of convergence. In Eq. (\ref{quantum_expander}), we collect terms of a
fixed power of $\beta$ when we use the  expansion of Eq. (\ref{s2}). In
the absence of a finite temperature phase transition, the series of Eq.
(\ref{quantum_expander}) converges for all $\beta$.  If $W$ vanishes on
a dense set of points along the real axis with arbitrary
then as $W$ is an analytic function then it must also vanish 
everywhere in the complex $\beta$-plane. All of the derivatives
of $W$ and, in
particular, its zeroth order value $W_{0}=[V, \hat{T}_{\mu}]$ must
vanish. However, $W_{0}$ is independent of the system in question (it
is independent of $H$, although $[H,\hat{T}_{\mu}]=0$). For any
non-trivial algebra captured by the operators $\{\hat{T}_{\mu}\}$,
there exists quasi-local operators $V$ that (i) have their support on
the same region where $\{\hat{T}_{\mu}\}$ are defined, and (ii) do not
commute with $\{\hat{T}_{\mu}\}$. Eq. (\ref{detectionT_cond+}) does not
preclude a commutator which vanishes in the thermodynamic limit. The
impossibility of Eq. (\ref{detectionT_cond}) in the thermodynamic limit
in all systems which exhibit SSB  detectable by quasi-local operators
follows from Eq. (\ref{VTV}). In all finite size systems, there is no
SSB (and the free energy is analytic - $\beta_{c} = \infty$)  and  Eq.
(\ref{detectionT_cond+}) cannot be satisfied at any non-zero
temperature. A similar conclusion holds for systems which, even in
their thermodynamic limit, do not exhibit $T>0$ transitions. Kitaev's
Toric code model is precisely such a system. It is important to
emphasize that the appearance of one (or more) $T>0$ transitions is a
necessary but not sufficient condition for Eq. (\ref{detectionT_cond})
to hold.

\section{Extensions of the Lieb-Robinson bound to non-local {\it
topological} quantities}
\label{LRS}

Locality in quantum spin systems is often well captured by a bound on
the commutator of local observables with disjoint supports. Early on,
Lieb and Robinson provided estimates and bounds on precisely such
commutators on nearest-neighbor spin systems on general graphs. These
naturally provide bounds for the speed of propagation of quantum
information. Additional works elaborated on the initial
findings of Lieb and Robinson and examined extensions to 
several bosonic and fermionic systems \cite{LR}. 

Stated formally, the Lieb-Robinson bound asserts that given two local
operators $V_{a}$ and $V_{b}$ in different regions ($a$ and $b$), the
operator norm of the commutator in gapped systems
\begin{eqnarray}
\| [V_{a}(0), V_{b}(t)] \|  \le A ({\sf vol}_{min}) \|V_{a}\|
\|V_{b}\|  e^{-(l- v|t|)/\xi}.
\label{LRE}
\end{eqnarray}
In Eq. (\ref{LRE}),  $l$ is the distance of the shortest path linking
regions $a$ and $b$ (more precisely the number of links on such a
path), ${\sf vol}_{min} = \min \{|a|, |b|\}$ denotes the number of
vertices in the smaller of the two sets $a$ and $b$. The constants
$A,v,$ and $\xi$ depend  on the maximal coordination of a given vertex
of the lattice and the largest local terms $\max\{\|h_{ij}\|\}$ which
appear in the argument of a nearest neighbor Hamiltonian system
\begin{eqnarray}
H = \sum_{\langle ij \rangle} h_{ij}.
\end{eqnarray}
Similar results are found to pertain to  finite temperature correlators
for systems  with local Hamiltonians. 

The extension of Eq. (\ref{LRE}) to non-local quantities is
straightforward. Let us consider the general product of local operators
over an $L^{d}$-dimensional volume $R_\mu$
\begin{eqnarray}
\hat{T}_{\mu} = \prod_{a \in R_\mu} V_{a}.
\end{eqnarray}
If the operator
norm of all $V$s is bounded from above by unity then 
the operator norm of the commutator
\begin{eqnarray}
&&\|[\hat{T}_{1}(0), \hat{T}_{2}(t)]\| \le  |R_{1}| |R_{2}| \max_{a' \in
R_{1},b' \in R_{2}}\|[V_{a'}(0), V_{b'}(t)]\| \nonumber \\ 
&&\hspace*{0.5cm} \le  A |R_{1}| |R_{2}| ({\sf vol}_{min}) \|V_{a'}\|~ \|V_{b'}\| 
e^{-(l_{\min}- v|t|)/\xi},
\label{ZL}
\end{eqnarray}
with $a' \in R_{1}$ and $b' \in R_{2}$. In Eq. (\ref{ZL}), $l_{\min}$
denotes the smallest distance between two points - one which belongs in
$R_{1}$ and the other in $R_{2}$. The first line of Eq. (\ref{ZL})
follows from a long-hand form of the expansion of the commutator. As
$\hat{T}_{1}$ contains $|R_{1}|$ terms and $\hat{T}_{2}$ has $|R_{2}|$
terms, there are a total of  $|R_{1}| |R_{2}|$ terms each containing a
string  of length $(|R_{1}| + |R_{2}| - 2)$ of local operators
$\{V_{a}\}$ which bracket a single commutator between two local
operators in the disjoint regions $[V_{a'}, V_{b'}]$. If the operator
norm of all $V$s is bounded from above by unity then  the bound of
the last line of Eq. (\ref{ZL}) follows from Eq. (\ref{LRE}).

As $l_{\min} \to \infty$, the equal time commutator between  two
distant {\it topological} operators vanishes.   In some cases, the
speed of propagation $v$ could be related to the velocity of defect
motion (e.g. domain walls). 

We can similarly bound the operator norm of the commutator which
appears in Eq. (\ref{detectionT_cond}). Here, we simply have that 
\begin{eqnarray}
\|[\hat{T}_{\mu}(0), V_b(t)]\| \le |R| \max_{a' \in
R_\mu,b}\|[V_{a'}(0), V_{b}(t)]\| \nonumber \\ 
\le  A |R| ({\sf vol}_{min}) \|V_{a'}\| ~\|V_{b}\|~  
e^{-(l_{\min}- v|t|)/\xi}.
\label{ZL+}
\end{eqnarray}

Although the influence of events far away from the support of
$\hat{T}_{\mu}$ (region $R_{1}$) can become nil (as seen from Eq.
(\ref{ZL+}), the cumulative effect of fluctuations on and proximate to
$R_{1}$ can (and indeed generally does) lead to a vanishing $\langle
\hat{T}_{\mu} \rangle$.

\section{The probability of observing anyons}
\label{anyp}

We now sketch how rigorous bounds on the probability of observing
anyons may be derived in one of the models that we discussed in this
work (Section \ref{kith}) -  Kitaev's Honeycomb model
\cite{Kitaev2006}.  Although some of the expressions given below are
special, the result seems to be of greater applicability. 

As shown by Kitaev, the GS configuration  is the anyon-free
configuration in which  $\langle \hat{O}_{h}\rangle= 1$  for all
plaquettes (hexagons) $h$. $\hat{O}_{h}$ is the product of the two
$\eta_{r}$ quantities which live on the center of two vertical bonds
which appear in $h$ \cite{CN}.   As (i) this system exhibits reflection
positivity (RP) \cite{RP} for the Ising-type free energy ($F_{\sf Ising}$) in the
argument of Eq. (\ref{Ztr}), 
\begin{eqnarray}
e^{-\beta F_{\sf Ising}[\{\eta_{r}\}]}  \equiv e^{-\frac{\beta}{2} \tr
[M_{\eta}]}  |\det (e^{\beta N_{\eta}} + 1)|^{1/2}
\label{GIsingW}
\end{eqnarray}
(ii) the minimum of the spectrum of  an $L \times L$ slab of the system
[of the square lattice of \cite{CN}] with open boundary conditions 
within a given topological sector $\{\hat{O}_{h}\}$ is discrete, and as
(iii) within the GS the system is anyon free, there must exist a length
$L_{c}$ such that in all slabs of size $L > L_{c}$ the lowest free energy
attainable over all states which are not anyon free is larger by a
finite gap $\Delta_{L}>0$ than the free energy within the  anyon
free infinite size system endowed with periodic boundary conditions.
As, for example, in Ref. \cite{bcn} and in particular \cite{zrp}, this
implies that the probability of finding anyons (or vortices in
\cite{zrp}) is exponentially suppressed at sufficiently low
temperatures. This is done by tiling  the lattice with all $L \times L$
blocks and noting that the probability for a given configuration is
bounded from above by 
\begin{eqnarray}
p = \exp \Big[ -\frac{\beta w N_{b} \Delta_{L}}{N_{c}} \Big] ,
\label{pRP}
\end{eqnarray}
where $w$ is a constant of order unity, $N_{b}$ is the number of {\it
bad} blocks which contain (at least) one anyon and $N_{c}$  is the
number of $L \times L$ blocks to which a given bond is common. Combined
with  RP of the Ising-type Gibbs weight of Eq. (\ref{GIsingW}), we can
prove that the probability of anyons is exponentially damped in
$\beta$. Thus, although a gap is generally required for {\it quantum
protection} from perturbations it may render the probability of
excitations of such anyons exponentially small. Equation (\ref{pRP}) is
a rigorous result for any uniform RP system  [including that of Eq.
(\ref{H})]. It is noteworthy that the introduction of {\em disorder}
may introduce anyons (and general defects) even at low  temperatures.
This has implications for impurities in  Quantum Hall systems.

\section{Thermal Equilibration time in systems with discrete symmetries} 
\label{equilta}

We can use the following heuristic {\it metastability} argument to
address the important question: How long does it take for a TQO system
with $d$-GLSs which has been prepared in the GS (protected subspace) to
equilibrate at a temperature $T$ in the presence of $d$-dimensional
defects? Stated alternatively, we now address the following question:
how long can the system retain  information before it is lost due to
ergodicity? The general answer to this question is quite complex. One
can use qualitative ideas borrowed from the theory of metastable
states, where the thermal equilibrium state is the global minimum of
the free energy $F$ while the GS manifold is the {\it suitable
subspace} which does not minimize $F$ but is supposed to be locally
stable. The lifetime of the protected subspace $\tau$ is clearly
determined by the dynamics of the system when coupled to a thermal
reservoir. The basic problem then reduces to characterizing the free
energy barrier $\Delta F$ that obstructs the escape from the GS
manifold. The crucial observation is that for a system of size $L
\times L \cdots \times L$ for a discrete symmetry of dimension $d>1$, 
the free energy barrier follows the form
\begin{eqnarray}
\Delta F = a L^{d-1} \sigma
\label{fa}
\end{eqnarray}
with $a$ a positive constant and  $\sigma$ the tension. The point is
that the presence of low-dimensional symmetries {\it lower} the height
of the barrier because of {\it dimensional reduction}. The case $d=1$
was considered in Section \ref{sec3} (although the energy cost was
finite in that case the entropy scaled logarithmically with the system
size). 

Let us assume that the probability to occupy an eigenstate
$\ket{\phi_\gamma}$ of $H$, 
\begin{equation}
P_\gamma(t)=\bra{\phi_\gamma} U^\dagger \rho U \ket{\phi_\gamma} ,
\end{equation}
with $\rho$ the initial state and $U$ a unitary evolution that includes
the coupling to the thermal bath, follows a master equation
\begin{equation}
\frac{d P_\gamma(t)}{d t} = \sum_{\gamma'\neq \gamma} (W_{\gamma \gamma'}
P_{\gamma'}(t) - W_{\gamma' \gamma}
P_{\gamma}(t)) ,
\label{master1}
\end{equation} 
with transition probabilities $W_{\gamma \gamma'}$ satisfying the
detailed balance condition (with equilibrium in a canonical ensemble)
\begin{equation}
W_{\gamma \gamma'} \exp[-\beta E_{\gamma'}]=W_{\gamma' \gamma}
\exp[-\beta E_{\gamma}] .
\label{master2}
\end{equation}
It is clear that the whole complexity of the dynamical problem is in
the functional form of the transition probabilities.  Let us assume
that there is only one possible channel for the system to escape from
the local minima.  

Kitaev's model is quite special in that it displays a $T=0$ critical
point. What can we say about systems that have a finite transition
temperature?

Typically, in systems with a finite critical temperature, the
equilibration time follows the Arrhenius form 
\begin{eqnarray}
\tau \sim \exp[\beta \Delta F]
\label{AR}
\end{eqnarray}
with $\Delta F$ the free energy barrier for the defects.  This, along
with Eq. (\ref{fa}) leads to the usual form 
\begin{eqnarray}
\tau \sim \exp[a \beta L^{d-1} \sigma].
\label{Ar}
\end{eqnarray}

In what follows, we address forms of this equilibration time and the
surface tension which governs it in several regimes.

\subsection{Equilibration time scaling near a critical point}

The considerations presented below apply general scaling arguments
(see e.g. \cite{glassme}) to general systems with discrete
symmetries. For such systems, next to the critical point,
\begin{eqnarray}
\sigma = k_{B} T_{c}/\xi^{d-1}
\end{eqnarray}
with $\xi$ the correlation length. Near the critical temperature $T_c$
the equilibration time scales as
\begin{eqnarray}
\tau \sim   \exp[c (L/\xi)^{d-1}]
\end{eqnarray}
with $c$ another positive constant. We can insert the scaling of $\xi$
with the reduced temperature $\overline{t}= (T-T_c)/T_c$ near $T_{c}$
\begin{eqnarray}
\xi \sim |\overline{t}|^{-\nu}
\end{eqnarray}
to obtain the relaxation time near the critical point
\begin{eqnarray}
\tau \sim A \exp[c L^{d-1} |\overline{t}|^{\nu (d-1)}].
\end{eqnarray}

\subsubsection{Finite surface tension when $T \to 0$}

In most systems (not all) having a finite $T_c$,  the surface tension
tends to a finite value as $T \to 0$. Consequently, Eqs. (\ref{fa},
\ref{AR}) imply that the equilibration time $\tau$ is strongly
divergent in both the system size $L$  (fixed temperature) and the
inverse temperature $\beta$ (at fixed $L$).

\subsubsection{Temperature dependent surface tension}

A temperature dependent surface tension may appear in  the orbital
compass model and other systems which  exhibit an order-out-of-disorder
phenomenon \cite{NBC,bcn}.  Here, we can prove that at sufficiently low
temperatures 
\begin{eqnarray}
\beta \sigma \ge q >0.
\end{eqnarray}
This, then, enables the proof of a finite-$T$ phase transition by the
usual Peierls argument. However, although the  relaxation time $\tau$
is divergent in system size it does not, for a finite fixed $L$,
diverge as  the temperature $T \to 0$.
%\vspace*{-2mm}

%\bibliography{gcs}

\begin{thebibliography}{28}
\bibitem{phase_trans}
In an SSB phase transition a Landau order parameter emerges at the
transition. There are other type of phase transitions where there is no
SSB of a global symmetry. However, in both cases, the existence of the
phase transition is signaled by a non-analyticity of their free
energies.
\bibitem{explain_stabilizer}
A working definition of the stabilizer of a code is the following. Let
$K$ be a subspace of an $n$-qubit Hilbert space. The stabilizer of $K$
is given by
\begin{eqnarray}
{\cal{S}}(K) = \{ P \in P_{n}| \ P|\psi \rangle = | \psi \rangle \mbox{
for all}~ | \psi \rangle \in K \}.
\end{eqnarray} 
Here, $P$ denote projection operators and $P_{n}$ is the projection
operator onto the $n$-qubit subspace $K$.
\bibitem{explain_14.5}
Wilson loops  can be expressed as
the products of the very same  plaquette operators that appear in the 
lattice stabilizer codes that are generalizations of gauge theory
Hamiltonians. [For instance, in the case of the Kitaev model which we
will discuss in Section \ref{d3kiv}, Wilson loops that circumscribe  a
region ${\sf D}$ can be expressed $W_{\partial {\sf D}} = \prod_{p \in
{\sf D}} B_{p}$ where the operator $B_{p}$ of Eq. (\ref{AB_defn}) was
invoked.] Here, condition (i) applies. As such, in lattice gauge theory
variants, the logical operators $\hat{T}_{\mu}$  cannot be taken as
Wilson loops and other general non-local operators $\hat{R}_{a}$.
\bibitem{explain_15.5}
If an expansion
about the (TQO) ordered  $T=0$ state is void - if the system exhibits a
$T=0$ transition - then $\langle  \hat{R}_{a} \rangle$ may vanish even
more rapidly as soon as temperature is made finite.  The inequality of 
Eq. (\ref{perimeter}) may be saturated (i.e. hold as an equality)
whenever an expansion about the  $T=0$ ordered state is possible. If no
low-$T$ ordered state exists and no expansion about the $T=0$ state is 
possible, then generally  $\langle \hat{R}_{a} \rangle_\alpha \to 0$
with the size of the spatial support of $\hat{R}_{a}$even faster 
then the "perimeter law" bound of Eq.(\ref{perimeter}). 
The simplest example is afforded by the two dimensional Ising 
gauge theory \cite{kogut} where for non-local
operators $\hat{R}_{a}$ which are Wilson loops, the expectation
values do not fall off as Eq.(\ref{perimeter}) but
rather by, generally, a much faster decay. Here, unless
the spatial support of $\hat{R}_{a}$ forms a closed loop,
$\langle \hat{R}_{a} \rangle =0$ identically by Elitzur's theorem.
For spatial supports of $\hat{R}_{a}$ that bound a given 
"Area",
we have that
\begin{eqnarray}
|\langle \hat{R}_{a} \rangle| \sim \exp[- \kappa \mbox{Area}] 
\label{WilsonloopA}
\end{eqnarray}
where $\kappa>0$ is a positive constant.
In the two dimensional $(D=2$) Ising gauge theory,
an expansion about the $T=0$ state is void.
Here, instead of Eq.(\ref{perimeter}), 
the much more dramatic fall-off of Eq.(\ref{WilsonloopA})
always holds. As before, $|\langle \hat{R}_{a} \rangle| \to 0$
in the limit of an infinite spatial support of
$\hat{R}_{a}$. 

Similarly, as we showed in \cite{NOLong}, 
Kitaev's Toric Code model exhibits a $T=0$
transition. In Kitaev's Toric Code model
an area law of the form of Eq.(\ref{WilsonloopA})
holds at all positive temperatures. Here, once
again, $|\langle \hat{R}_{a} \rangle| \to 0$
in the limit of an infinite spatial support of
$\hat{R}_{a}$.
\bibitem{explain_15.75}
In lattice gauge theories in which
such $d=0$ are realized, all Wilson loops \cite{kogut} are invariant
under all local (i.e. $d=0$) symmetries and may thus attain a finite
expectation value at positive temperatures. However, as noted above
Wilson loops are part of the code's stabilizer. A Wilson loop is
related to the product of all of the plaquette variables of all
plaquettes inside the loop. Wilson loops and all local symmetries
further commute with one another and do not realize the non-trivial
algebra necessary for topological degeneracy: the operators $\{
\hat{T}_{\mu} \}$  are not Wilson loops. The absence of a SSB of
$\hat{T}_{\mu}$ detectable by any local measurement is the physical
content of Eq. (\ref{VTVT}).
\end{thebibliography}

\end{document}